\documentclass{aa}  
\usepackage{graphicx}
\usepackage{txfonts}
\usepackage{aalongtable}
\usepackage{natbib}
\bibpunct{(}{)}{;}{a}{}{,}             
\begin{document} 

 \title{An imaging and spectroscopic study of the planetary nebulae in 
NGC~5128 (Centaurus~A)\thanks{Based on observations collected at the ESO La Silla-Paranal 
Observatory within observing programmes 060.A-9140 (FLAMES Science Verification) 
and 073.B-0434.}}

   \subtitle{Planetary nebulae catalogues\footnote{Tables 7 -- 11 are only available in electronic form
at the CDS via anonymous ftp to cdsarc.u-strasbg.fr (130.79.128.5)
or via http://cdsweb.u-strasbg.fr/cgi-bin/qcat?J/A+A/}
}

   \author{J. R. Walsh\inst{1}
          \and 
          M. Rejkuba\inst{1,2}
          \and
          N. A. Walton\inst{3}
          }

   \institute{European Southern Observatory, Karl-Schwarzschild Strasse 2,
              D-85748 Garching, Germany \\
              \email{[jwalsh], [mrejkuba]@eso.org}
	 \and
          Excellence Cluster Universe, Boltzmannstr.\ 2, D-85748, Garching, Germany 	 
         \and
             Institute of Astronomy, University of Cambridge,
             Madingley Road, Cambridge CB3 OHJ, United Kingdom
             \email{naw@ast.cam.ac.uk}
             }

   \date{Received 30/05/2104; accepted 08/09/2014}
   \titlerunning{Imaging and kinematic study of PNe in NGC~5128}
   
 
  \abstract
{Planetary nebulae (PNe) are excellent tracers of the common low mass 
stars through their strong and narrow emission lines. The
velocities of large numbers of PNe are excellent tracers of 
galaxy kinematics. NGC~5128, the nearest large early-type galaxy,  
offers the possibility to gather a large sample.}
{Imaging and spectroscopic observations 
of PNe in NGC~5128 were obtained to find and measure their velocities. 
Combined with literature data, a large sample of high quality kinematic 
probes is assembled for dynamical studies.}
{NTT imaging was obtained in 15 fields in NGC~5128 across 1$^\circ$ 
with EMMI and [O~III] and off-band filters.
Newly detected sources, combined with literature PN, 
were used as input for FLAMES multi-fibre 
spectroscopy in MEDUSA mode. Spectra of the 
4600-5100\AA\ region were analysed and velocities measured 
based on [O~III]4959,5007\AA\ and often H$\beta$.}
{The chief results are catalogues of 1118 PN candidates and 1267
spectroscopically confirmed PNe in NGC~5128. The catalogue of PN
candidates contains 1060 PNe discovered with NTT EMMI imaging
and 58 from literature surveys. The spectroscopic PN 
catalogue has FLAMES radial velocity and emission line measurements for 
1135 PNe, of which 486 are new. Another 132 PN
radial velocities are available from the literature. For 629 PNe 
observed with FLAMES, H$\beta$ was measured in addition to [O~III]. 
Nine targets show double-lined or more complex profiles, and their 
possible origin is discussed. FLAMES 
spectra of 48 globular clusters were also targetted: 11 had emission 
lines detected (two with multiple components), but only 3 are PNe likely 
to belong to the host globular. 
}
{The total of 1267 confirmed PNe in NGC~5128 with radial velocity 
measurements (1135 with small velocity errors) is the largest collection of 
individual kinematic probes in an early-type galaxy. This 
PN dataset, as well as the catalogue of PN candidates, are 
valuable resources for detailed investigation of NGC~5128.}

\maketitle
%

\section{Introduction}
NGC~5128 is the closest example of a large early-type (gE/S0) galaxy and it has been 
the subject of extensive studies. Many of these have concentrated on its 
active nucleus, which is observed over the whole electromagnetic spectrum 
from $\gamma$ rays to low frequency radio \citep[see][for an earlier review]{israel98}, 
and the galaxy is often referred to by its radio source name as Centaurus A. 
NGC~5128 shows signs of major nuclear activity, including the AGN and 
supermassive black hole \citep[][ and references therein]{neumayer10}.
There is also evidence of past interaction events, in the form of a dust lane 
with a ring of young clusters and HII regions \citep{kainulainen+09, graham79}, 
a young ($\sim$0.3 Gyr) tidal stream \citep{peng+02} and stellar shells in its outer 
regions \citep{malin+83}. However despite its disturbed past, it may well be a 
typical early-type galaxy dominated by an old stellar population.
%
The Sauron 
and ATLAS$^{3D}$ surveys have shown that at least 40\% of local early-type galaxies 
show signs of interactions with HI, molecular emission
\citep{combes+07, davis+11,young+13}, extended emission line gas \citep{sarzi+07}
and multiple kinematic components. \citet{harris10} provides the most recent review of the 
underlying elliptical galaxy properties, including its large population of globular 
clusters and planetary nebulae.

NGC~5128 shows signs of major nuclear activity, including the AGN and 
supermassive black hole \citep[][ and references therein]{neumayer10}.
There is also evidence of past interaction events, in the form of a dust lane 
with a ring of young clusters and HII regions \citep{kainulainen+09, graham79}, 
a young ($\sim$0.3 Gyr) tidal stream \citep{peng+02} and stellar shells in its outer 
regions \citep{malin+83}. However despite its disturbed past, it may well be a 
typical early-type galaxy dominated by an old stellar population.

The first extensive catalogue of planetary nebulae (PNe) in NGC~5128 
was released by \citet{hui+93b} using the now classical 
technique of difference filter imaging on the strong [O~III]5007\AA\ 
emission line and on a redder band that includes only very weak emission
lines. The 785 PN candidates were catalogued. From their [O~III] magnitudes, 
\citet{hui+93a} determined a distance of 3.5 Mpc to NGC~5128 by 
fitting the planetary nebula luminosity function (PNLF) and comparing 
it to that of M31. Independent measurements from the Mira 
period-luminosity relation and the luminosity of the tip of the red 
giant branch \citep{rejkuba04, rejkuba+05}, 
surface brightness fluctuations \citep{tonry+01}, globular 
cluster luminosity function \citep{harris+88} and 42 
classical Cepheid variables \citep{ferrarese+07}
have resulted in distance estimates in the range 3.4 to 4.1 Mpc. 
\citet{harris+10} present a comprehensive review 
of distance estimates to NGC~5128 and recommend a value
of 3.8 Mpc, which is adopted here (distance modulus $27.91 \pm 0.08$ mag). 

The same PN detection technique was employed by \citet{peng+04PN} 
to extend the initial detection survey of \citet{hui+93b} to an additional set of 
356 PN candidates, concentrating on the outer regions and reaching as far as 
$\sim 15 r_e$ (87 kpc for distance corrected to 3.8 Mpc) and 
including four PNe at radii $>$65 kpc. Thus together with the 785 PN candidates 
from \citet{hui+93b}, these two surveys have brought the 
total number of PN candidates in NGC~5128 to 1141\footnote{Note that 
\citet{peng+04PN} mention that one pair of PN candidates from \citet{hui+93b} 
(also contained in \citet{hui+95}) is actually the same object, and therefore the total number should
have been reported as 1140. We have made careful checks of duplicate entries 
and discuss astrometry in section~\ref{sec:astrometry}.}, comparable to the 
number known in the Milky Way and M~31  \citep{parker+12, merrett+06}, 
making it the third largest galaxy PN sample to date.

The confirmation of PN candidates from imaging surveys as bona fide 
PNe is usually performed by spectroscopic detection of the narrow emission lines, 
particularly of the [O~III] doublet at 4959, 5007\AA. The detection 
efficiency of the on/off-band imaging technique is generally high. 
This arises since the [O~III]5007\AA\ line has a large equivalent 
width, the nebular continuum is low and the star is hot, so the 
contrast of the on- and off-band images is large. Aside from the 
false detections through poorly matched photometry or on- and off-band 
[O~III] images taken under different observing conditions, potential 
contaminants are compact HII regions (such as are present in the vicinity
of the dust lane of NGC 5128), symbiotics stars, supernova remnants 
and emission line galaxies at higher 
redshift, such as [O~II] emitters at z=0.34 or even Lyman-$\alpha$ 
emitters at z=3.1 \citep[e.g.\ see][]{kudritzki+99}. 

Early spectroscopy of PNe in NGC~5128 was performed by \citet{Walsh+99} who
observed five PNe with long slits and analysed their reddening and 
O/H abundance ratios. A larger sample of PNe, selected from the catalogue of 
\citet{hui+93b}, was observed with FORS multi-slits
and spectra of 51 PNe were measured in three fields at offsets from
4 to 17 kpc (\citet{Walsh+12}). Fainter diagnostic lines for
electron temperature and density determination were measured in
a fraction of these PNe. All targets showed entirely characteristic
PN spectra and light element abundances (He, N, O, Ne)
were determined by comparison to photoionization models. A mean
[O/H] of -0.17 was determined, with no evidence of a radial gradient 
in O/H.

Other spectroscopic studies have focussed on PNe as kinematic probes. 
\citet{hui+95} obtained multi-object (fibre-linked) spectroscopy of some 
of their PN candidates and measured radial 
velocities from the [O~III]5007\AA\ line for 433 (corrected to 431 unique PNe
 by \citet{peng+04PN}, after accounting for repeat detections). Two spectrometers
were employed giving resolutions of 2 and 4\AA\ and 1$\sigma$ radial
velocity errors of 4 and 30 kms$^{-1}$ respectively. Out to
10 kpc, the PN velocity field was found to have features of a triaxial
potential, such as an offset between the rotation axis and the minor axis
and the line of maximum rotation. The isotropic Jeans equation was applied
to the PN position and kinematic data providing evidence of an increased 
mass-to-light ratio with radius consistent with the presence of a dark 
matter halo. 

\citet{peng+04PN} targeted spectroscopically  353 PN candidates from \citet{hui+93b}, 
not confirmed previously by \citet{hui+95}, as well as their own 356 PN candidates. 
They used AAO/2dF and CTIO ARGUS and Hydra spectrometers which
provided resolutions from 2.2 to 4.9\AA. The spectral range was centred on [O~III]5007\AA\ 
except for two fields along the minor axis that included H$\alpha$ instead. 
Combining with the literature data \citep{hui+95} and removing duplicates, \citet{peng+04PN} 
presented a catalogue of 780 spectroscopically confirmed PN in NGC~5128, adding 
349 newly confirmed PN from their spectroscopic follow-up. The
more extensive spatial coverage and number of velocity probes led to
more detail on the velocity field, such as the large rotation along the
galaxy major axis and a pronounced twist of the zero-velocity contour,
indicative of a triaxial potential, probably prolate. The total dynamical
mass of the galaxy, including the dark matter halo, was measured to
be $\sim 6 \times 10^{11}$ M$_\odot$, but the mass-to-light ratio 
($M/L_{B}$) was estimated to be abnormally low with respect to other elliptical 
galaxies. 

Another source of kinematic probes in NGC~5128 are the extensive
catalogue of globulars clusters (GCs). Currently there are 564 
GCs with confirmed radial velocity measurements 
\citep{VHH81, hesser+84, hesser+86, harris+92, peng+04GCcat, woodley+05,
woodley+10a, woodley+10b, rejkuba+07,  beasley+08}.
\citet{woodley+07} presented a combined kinematic study based on 780 PN velocities 
\citep{peng+04PN} and 340 GC velocities.
There are some differences between the PN and GC kinematics such as
higher rotation amplitude of 76 kms$^{-1}$ for PN compared to 47 and
31 kms$^{-1}$ for the metal-rich (red) and metal poor (blue) globulars, respectively.
Using the tracer mass estimator, as done by \citet{peng+04PN}, 
the mass supported by rotation is $0.9 \times 10^{12}$M$_\odot$ from the PN data 
while for the combined GC data set, the total mass is estimated at $1.2 \times 10^{12}$M$_\odot$,
both consistent within the errors. \citet{harris+12} have 
presented a list of 833 new high-quality GC candidates 0.8 mag fainter than the 
globular cluster luminosity function turnover point, which will provide a further 
rich source for kinematic studies, after acquiring necessary radial velocity data.

\citet{woodley+11} used both the position and velocity
data on the PNe and GCs in NGC~5128 to search for possible sub-groups 
that may trace satellite galaxy accretion events onto the giant elliptical host galaxy.
Using a position and kinematic search criterion, four subgroups based on PNe 
and four based on GCs were localised with two of the subgroups in common.
While the role of both GCs and PNe as kinematic probes of a galaxy is 
highly complementary, and for a full picture of the global velocity field
both are obviously required, the emission line nature of the PN spectra offers
some advantages, in particular ease of spectroscopic confirmation and
simplicity and accuracy of measurement of radial velocities (no templates 
are required and the emission lines are narrow).

With the advent of the FLAMES multi-object facility at the ESO VLT, it
became possible to extend the 4 m class telescope PN kinematic 
surveys of \citet{hui+95} and \citet{peng+04PN} with 
higher spectral resolution, leading to some distinct advantages. 
The higher throughput allows the 3-times weaker 4959\AA\
line of the [O~III] doublet to be detected for all but the
faintest PNe, thus breaking any possible degeneracy through the 
detection of only a single line and leading to better discrimination 
against higher redshift emission line sources, together with improved velocity 
determinations based on multiple lines. Also the 
possibility to detect the H$\beta$ line helps to confirm the PN nature
of the sources, distinguishing them, for example, from compact HII regions and X-ray nebulae
associated with binary sources. The higher spectral resolution allows
high absolute velocity accuracy to be attained, reducing the uncertainties
on the velocity dispersion of the sample, with concomitant advantages for 
the velocity modeling. The larger telescope aperture of course enables the radial 
velocity of fainter PNe to be recorded since they are point sources at the
distance of NGC~5128, thus increasing the sample. 


The effectiveness of the FLAMES GIRAFFE spectrograph for 
PN spectroscopy was first tested during the science verification (SV) 
of the instrument in 2003.  The match between the instrument 
multiplex and the sensitivity showed that at least the strongest 
[O~III]5007\AA\ line can be detected in all PN found from previous imaging 
surveys. We thus extended these imaging surveys by
undertaking a survey with the 3.58 m NTT in La Silla using EMMI instrument to detect 
more PNe across the surface of NGC~5128. The resulting PN candidates, complemented
with additional PN candidates from the catalogue of \citet{peng+04PN} (that includes
also the \citet{hui+93b} PN candidates) were observed in 10 FLAMES pointings (diameter $\sim$25$'$) 
within a service mode programme in the ESO Period 73. This paper describes these observations
and the resulting new catalogue, that almost doubled the number of previously 
known PN with spectroscopically measured velocities in NGC~5128.

The imaging observations with EMMI and the reductions are described in 
section~\ref{sec:imaging}. Section~\ref{sec:spectroscopy} presents all the FLAMES 
observations from both SV and the later ESO Period 73 proposal. 
Section~\ref{sec:rv} is devoted to the reduction of
the GIRAFFE spectra and derivation of the radial velocities and presentation
of the catalogue of kinematic data. In section~\ref{sec:catalogs} the catalogues
of new PN candidates and spectroscopically confirmed PNe are presented. The total 
number of PNe with a velocity determination in NGC~5128 now reaches 1267. 
Section~\ref{sec:results} 
discusses various aspects of the data such as PN spectra with unusual velocity structure, 
emission line sources at the positions of globular clusters and comparison with 
X-ray source catalogues. The discussion briefly summarizes aspects of the data and 
outlines potential follow-on studies. The data in the form of PN position, radial 
velocity and, where available, an [O~III]/H$\beta$ line ratio, are made 
available electronically at CDS, and we encourage exploitation of this high quality 
data set.

\section{Imaging data}
\label{sec:imaging}

\subsection{EMMI Imaging Observations}

\begin{table}
\caption{EMMI imaging filters}
\begin{tabular}{lccc}
\hline\hline
\multicolumn{1}{c}{Filter}           &  Central $\lambda$ &  Width &   Peak T \\
\multicolumn{1}{c}{Name}             &   (nm)             & (nm)   &    (\%)   \\
\hline
OIII\#589        &  501.1  &   5.5  & 61 \\
 Spe\#767        &  546.4  &  20.7  & 75 \\
H$\alpha$\#596   &  656.8  &   7.3  & 54  \\
H$\alpha$\#601   &  689.4  &   7.3  & 58  \\
\hline
\end{tabular}
\label{tab:filters}
\end{table}

\begin{figure}
\centering
\resizebox{\hsize}{!}{
\includegraphics[angle=0,clip]{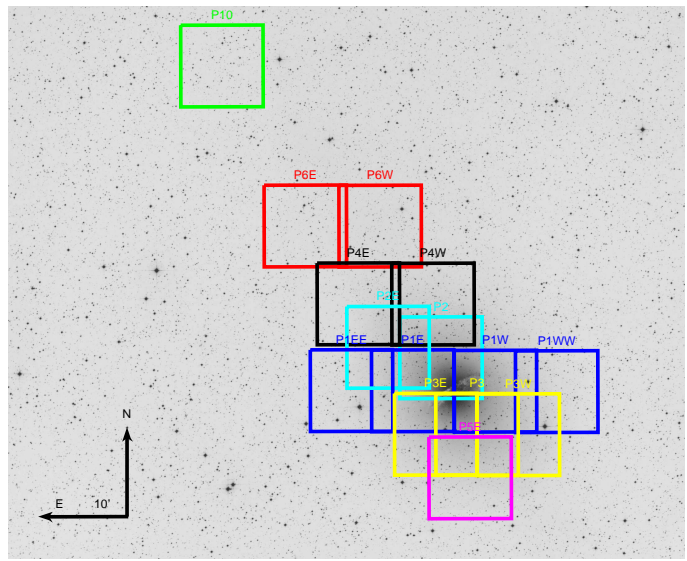}
}
\caption{Distributions of the fields imaged with EMMI on the ESO NTT are shown on the
DSS2 red image of NGC~5128.}
\label{fig:EMMI_fields}
\end{figure}

The imaging observations were carried out with the ESO Multi Mode Instrument
\citep[EMMI; ][]{emmi} mounted on the Nasmyth  B focus of the ESO New  
Technology Telescope (NTT) during 4 visitor mode observing 
nights 12-16 May 2004 allocated to the observing run 073.B-0434(A). 

EMMI is equipped with two $2\mathrm{k} \times 4\mathrm{k}$ MIT/LL CCDs 
with $15 \times 15$ $\mu$m pixels, 
yielding $0\farcs 166$ pixel size and the field size of $9\farcm 9 \times 9\farcm 1$. 
The two CCDs are mounted next to each other with a gap of 47 pix ($7\farcs82$) in between. 
The EMMI images are in Multi-Extension Format (MEF) with 4 extensions, corresponding 
to the $2\times2$ amplifiers. However, the size of the image in the extension 4 is 
very small (with useful data recorded over less than 14" on a side), the number 
of sources detected there was negligible and some were spurious. Therefore, 
in the following it was decided to use only sources detected in the first three 
extensions (an area of $20 \farcm 84^{2}$).

The plan was to cover the $20 \times 10$~kpc$^2$ (major $\times$ minor axis) 
with 22 fields oriented along the major axis (PA=35$^\circ$) using EMMI 
red arm imaging mode. Each field was planned to receive 80 min exposure with 
[O~III] exposures taking 60 min and the off-band continuum filter 20 min, with 
the aim to reach at $3\sigma$ PN candidates with m$_{5007 \AA} = 27.5$. The 
on-band filter was [O~III]\#589, centred at 5011\AA\  and the off-band filter 
was the so-called SPE\#767 centred at 5464\AA. Basic characteristics of the 
filters are given in Table~\ref{tab:filters}.

\begin{table*}
\caption[]{Log of NTT EMMI on- / off-band imaging }
\label{tab:EMMI_log}
\centering 
\begin{tabular}{l l l l r c l l l} 
\hline\hline  
 Field & ~~~~~~~$\alpha$ & ~~~~~~~$\delta$ & UT date & T$_{exp}$ & Filter & Airmass & IQ$^{\ast}$ & Note \\
         &  (h ~~m ~s) & ($^\circ$ ~~~$\arcmin$ ~~~$\arcsec$)    &  &  (sec)  &  &  & ($''$)  & \\
\hline  \\
 NGC5128-P1 & 13 25 28 & -43 01 10 & 2004-05-13 & 2x1800 & OIII\#589 & 1.39 & 0.75 & \\
 NGC5128-P1 & 13 25 28 & -43 01 10 & 2004-05-13 & 2x600 & SPE\#767 & 1.21 & 0.83 & \\

 NGC5128-P1 & 13 25 28 & -43 01 10 & 2004-05-14 & 2x1200 & HA\#596 & 1.03 & 0.65 & \\
 NGC5128-P1 & 13 25 28 & -43 01 10 & 2004-05-14 & 2x1200 & HA\#601 & 1.03 & 0.60   & \\
 NGC5128-P1 & 13 25 28 & -43 01 10 & 2004-05-14 & 1x600 & SPE\#767 & 1.45 & 0.60   & \\

 NGC5128-P1E & 13 25 53 & -43 01 10 & 2004-05-14 & 2x1800 & OIII\#589 & 1.4 & 0.75 & \\
 NGC5128-P1E & 13 25 53 & -43 01 10 & 2004-05-14 & 2x600 & SPE\#767 & 1.06 & 0.60 & \\

 NGC5128-P1EE & 13 26 30 & -43 01 10 & 2004-05-14 & 2x1800 & OIII\#589 & 1.06 & 0.60  & \\
 NGC5128-P1EE & 13 26 30 & -43 01 10 & 2004-05-14 & 2x600 & SPE\#767 & 1.13 & 0.55 & \\

 NGC5128-P1W & 13 25 03 & -43 01 10 & 2004-05-14 & 2x1800 & OIII\#589 & 1.20 & 0.58 & \\
 NGC5128-P1W & 13 25 03 & -43 01 10 & 2004-05-14 & 2x600 & SPE\#767 & 1.37 & 0.58  &\\

 NGC5128-P1WW & 13 24 26 & -43 01 10 & 2004-05-14 & 2x1800 & OIII\#589 & 1.62 & 0.60 & \\
 NGC5128-P1WW & 13 24 26 & -43 01 10 & 2004-05-14 & 2x600 & SPE\#767 & 2.02 & 0.75 & \\

 NGC5128-P2 & 13 25 36 & -42 57 29 & 2004-05-12 & 2x1800 & OIII\#589 & 1.27 & 0.60     & Elliptical images \\
 NGC5128-P2 & 13 25 36 & -42 57 29 & 2004-05-12 & 1x300  & SPE\#767 & 1.44 & 0.56  &  \\
 NGC5128-P2 & 13 25 36 & -42 57 29 & 2004-05-12 & 2x600  & SPE\#767 & 1.381 & 0.65 & \\

 NGC5128-P2E & 13 26 08 & -42 56 20 & 2004-05-13 & 3x1800 & OIII\#589 & 1.12 & 0.45 & Elliptical images\\
 NGC5128-P2E & 13 26 08 & -42 56 20 & 2004-05-16 & 2x1800 & OIII\#589 & 1.11 & 0.65  & \\ 
 NGC5128-P2E & 13 26 08 & -42 56 20 & 2004-05-13 & 2x600 & SPE\#767 & 1.05 & 0.50   & Elliptical images \\
 
 NGC5128-P2E & 13 26 08 & -42 56 20 & 2004-05-13 &  2x600 &  SPE\#767& 1.03 & 0.44 & \\
NGC5128-P2W & 13 25 19 & -42 56 20 & 2004-05-13 & 2x1800 & OIII\#589 & 1.17 & 0.47 & \\
 
 NGC5128-P3 & 13 25 14 & -43 06 00 & 2004-05-13 & 2x1800 & OIII\#589 & 1.05 & 0.50 & \\
 NGC5128-P3 & 13 25 14 & -43 06 00 & 2004-05-13 & 2x600 & SPE\#767 & 1.1 & 0.45 & \\

 NGC5128-P3E & 13 25 39 & -43 06 00 & 2004-05-13 & 2x1800 & OIII\#589 & 1.45 & 0.55 & \\
 NGC5128-P3E & 13 25 39 & -43 06 00 & 2004-05-13 & 2x600 & SPE\#767 & 1.31 & 0.50   & \\

 NGC5128-P3W & 13 24 49 & -43 06 00 & 2004-05-15 & 2x1800 & OIII\#589 & 1.12 & 0.45 & \\
 NGC5128-P3W & 13 24 49 & -43 06 00 & 2004-05-15 & 2x600 & SPE\#767 & 1.07 & 0.40  & \\

 NGC5128-P4E & 13 26 26 & -42 51 30 & 2004-05-15 & 2x1800 & OIII\#589 & 1.17 & 0.50  &  \\
 NGC5128-P4E & 13 26 26 & -42 51 30 & 2004-05-15 & 2x600 & SPE\#767 & 1.07 & 0.50 &  \\

 NGC5128-P4W & 13 25 41 & -42 51 30 & 2004-05-15 & 3x1800 & OIII\#589 & 1.03 & 0.43 &  \\
 NGC5128-P4W & 13 25 41 & -42 51 30 & 2004-05-15 & 2x600 & SPE\#767 & 1.05 & 0.35 &  \\

 NGC5128-P5E & 13 25 18 & -43 10 50 & 2004-05-15 & 2x1800 & OIII\#589 & 1.27 & 0.54  & \\
 NGC5128-P5E & 13 25 18 & -43 10 50 & 2004-05-15 & 2x600 & SPE\#767 & 1.47 & 0.55  &  \\

 NGC5128-P5W & 13 24 29 & -43 10 50 & 2004-05-15 & 2x1800 & OIII\#589 & 1.75 & 0.75  & \\
 NGC5128-P5W & 13 24 29 & -43 10 50 & 2004-05-15 & 2x600 & SPE\#767 & 2.28 & 1.05 & \\

 NGC5128-P6E & 13 26 58 & -42 42 50 & 2004-05-16 & 3x1800 & OIII\#589 & 1.03 & 0.50  & Elliptical \\
 NGC5128-P6E & 13 26 58 & -42 42 50 & 2004-05-16 & 2x600 & SPE\#767 & 1.06 & 0.45 &  \\

 NGC5128-P6W & 13 26 13 & -42 42 50 & 2004-05-16 & 2x1800 & OIII\#589 & 1.24 & 0.55  &  \\
 NGC5128-P6W & 13 26 13 & -42 42 50 & 2004-05-16 & 2x600 & SPE\#767 & 1.44 & 0.56  &  \\

 NGC5128-P7E & 13 24 42 & -43 19 15 & 2004-05-16 & 1x1800 & OIII\#589 & 1.83 & 0.63  & Thin clouds\\
 NGC5128-P7E & 13 24 42 & -43 19 15 & 2004-05-16 & 1x1000 & OIII\#589 & 1.83 & 0.76  & Thin clouds\\
 NGC5128-P7E & 13 24 42 & -43 19 15 & 2004-05-16 & 2x600 & SPE\#767 & 2.24 & 1.10  & Thick clouds! Guiding problems\\

 NGC5128-P10 & 13 27 48 & -42 25 00 & 2004-05-13 & 1x1800 & OIII\#589 & 1.78 & 0.45 & \\
 NGC5128-P10 & 13 27 48 & -42 25 00 & 2004-05-13 & 1x600 & SPE\#767 & 2.10 & 0.68  & \\

 NGC5128-P10 & 13 27 48 & -42 25 00 & 2004-05-14 & 1x1800 & OIII\#589 & 1.34 & 0.45 &  \\
 NGC5128-P10 & 13 27 48 & -42 25 00 & 2004-05-14 & 1x600 & SPE\#767 & 1.24 & 0.88  & \\
\hline
\end{tabular}
\tablefoot{
$^{\ast}$ IQ is the image quality (FWHM in arcseconds) measured on the images.
}
\end{table*}

The sky was clear or photometric for the first three nights, with the last 
night having thin cirrus in the beginning and then thick clouds at the end. 
Seeing was generally excellent, resulting in 2.5--4.5 pix ($0\farcs42 - 0\farcs75 $) FWHM 
images, except for the end of the third and fourth nights, when it degraded.
As a result, bad image quality and low limiting flux occurred for the F7E 
field off-band images taken on the fourth night. There were too many detections 
in the [O~III] images of this field that had no counterpart in the off-band images 
due to thick clouds. Similar problems were also encountered with the P5W field for which 
data were taken at high airmass at the end of third night. Therefore it was decided not 
to follow-up these fields further. Field P1 overlaps 
completely with P1W and P1E and therefore sources from these fields are in common. 
$H\alpha$ images for field P1 were also obtained, with the aim to distinguish between 
potential H\,II regions in the central parts of the galaxy and PN candidates, but these 
images are not discussed here. Hence, finally here we concentrate on the 15 
fields with best quality data for PN candidate search. The distribution of these 
fields is shown in Fig.~\ref{fig:EMMI_fields}, while for completeness
all EMMI observations are listed in the observing log in Table~\ref{tab:EMMI_log};
all coordinates are J2000.

\subsection{EMMI data reduction}

\begin{table}
\caption{Off-band, [O~III] on-band and PN candidate detections in EMMI images.}
\begin{tabular}{lrrr}
\hline \hline
\multicolumn{1}{c}{Field}  &\multicolumn{1}{c}{Off-band}  & \multicolumn{1}{c}{[O~III]}  & \multicolumn{1}{c}{PN}    \\
\multicolumn{1}{c}{ID} & \multicolumn{1}{c}{detections}   & \multicolumn{1}{c}{detections} & \multicolumn{1}{c}{candidates} \\
\hline
NGC5128-P1E	&	1415	 &	1183	 &	165	\\
NGC5128-P1EE &	1184	 &	1033 &	92	\\
NGC5128-P1W &	1288	 & 	1241	 &	212	\\
NGC5128-P1WW &	1061	 &	1179 &	133	\\
NGC5128-P2 &	1180 &	1421 &	333	\\
NGC5128-P2E &	1658	 &	1233 &	191	\\
NGC5128-P3 &	1611	 &	1521 &	282	\\
NGC5128-P3E &	1586 &	1394 &	224 \\
NGC5128-P3W &	1783 &	1507 &	237	\\
NGC5128-P4E	 &	1717	 & 	1505	 &	170	\\
NGC5128-P4W &	2111 &	1751	 & 	218	\\
NGC5128-P5E	 &	1515	 & 	1290	 & 	110	\\
NGC5128-P6E	 &	1078 &	1029	 &	59	\\
NGC5128-P6W &	1348	 &	1205 &	93	\\
NGC5128-P10	 &	1668 &	1842	 &	42	\\
\hline
Total                    &                &                &    2561      \\
\hline
\end{tabular}
\label{tab:emmi_detections}
\end{table}

The primary goal of the EMMI imaging run 073.B-0434(A) was to identify PN candidates
for the spectroscopic follow-up with FLAMES run 073.B-0434(B). To speed up the
preparation of the already scheduled follow-up spectroscopic observations, we used 
only the first [O~III] and the first off-band exposures to create the list of PN 
candidates: thus the detection images had 1800sec exposure for 
on-band and 600sec for the off-band data. Data reduction was done
using the Cambridge Astronomy Survey Unit (CASU) pipeline and consisted of bias
subtraction, flat-fielding, astrometric correction and source extraction using 
aperture photometry.  

The on- and off-band detections were then matched and the PN candidates were 
identified with on-band sources that did not have an off-band counterpart. 
The number of sources detected in the [O~III] and off-band filters and the 
number of PN candidates in each field are listed in 
Table~\ref{tab:emmi_detections}. Combining the 
two exposures per filter should result in a more complete catalogue of PN 
candidates in NGC~5128, yielding some additional fainter sources, 
and is being undertaken. To assess
the quality and reliability of these fainter sources extensive completeness
simulations and possibly some spectroscopic follow-up will need to be done. 

The 2561 EMMI PN candidate sources were first culled from double detections due to
field overlaps. The 401 groups of objects  (280 pairs and 121 triplets) were identified within a $0\farcs5$ 
circle on the sky, hence decreasing the EMMI PN candidate list to 2039 unique sources
\footnote{Within 1" circle there are 410 groups (287 pairs and 123 triplets),  which 
implies 2028 unique sources within that slightly larger matching radius.}. 
The candidate list was matched with PN catalogues from the literature in order to 
build a comprehensive catalogue of PN candidates.

\subsection{Astrometry and catalogue matching}
\label{sec:astrometry}

Comparing the positions of 280 matched pairs in overlapping EMMI images, a
$1\sigma$ internal astrometry error of $0\farcs15$ was estimated for the EMMI 
imaging catalogue. The external accuracy was estimated by matching the EMMI 
imaging catalogue with the \citet{peng+04PN} catalogues of PN candidates and 
spectroscopically confirmed PN, as well as with their catalogue of stars (tables 6, 7 and 3 in  
\citet{peng+04PN}, respectively). The latter yielded only one star within $2\farcs 0$ 
in RA and DEC from the PN candidate sources, giving a high confidence that most of 
the sources that we detect in [OIII], but not in the off-band image, are indeed bona fide PN. 

\begin{figure}
\centering
\resizebox{\hsize}{!}{
\includegraphics[angle=0]{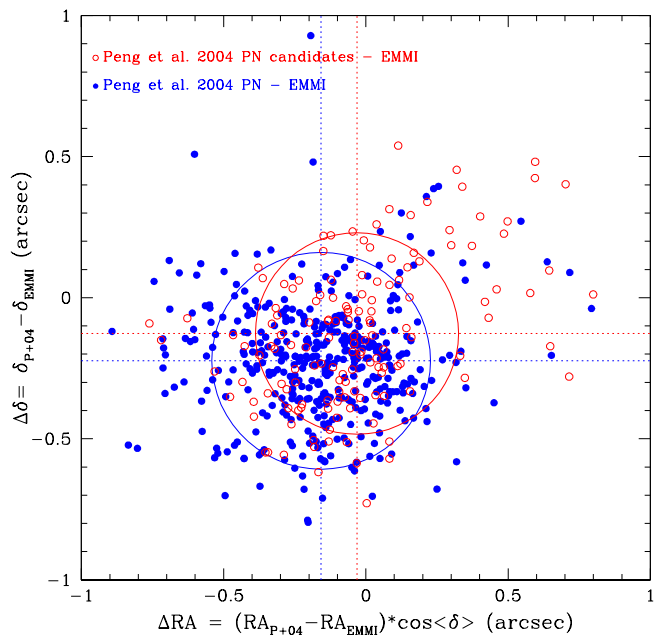}
}
\caption{Coordinate difference between EMMI and \citet{peng+04PN} PN candidates (open circles)
and PN with confirmed radial velocities (filled dots). The circles are the mean separation for all sources 
matched within 1". For  PN candidates the mean separation is $0\farcs 36 \pm 0\farcs 19$, similar to 
$0 \farcs 38 \pm 0\farcs 19$ for confirmed PN. }
\label{fig:radecEMMI_Peng}
\end{figure}

\begin{table}
\caption{List of pairs of PN candidates from the Table 7 of \citet{peng+04PN} located within $<1"$ on the sky.}
\begin{tabular}{llc}
\hline\hline
  \multicolumn{1}{c}{Name} &
  \multicolumn{1}{c}{Name} &
  \multicolumn{1}{c}{Separation} \\
  \multicolumn{1}{c}{(Hui et al.1993b)} &
  \multicolumn{1}{c}{(Peng et al.2004)} &
  \multicolumn{1}{c}{(")} \\
  \hline
  509 & f14p021 & 0.20\\
  511 & f14p030 & 0.41\\
  548 & f14p036 & 0.32\\
  4201 & f14p001 & 0.55\\
  4287 & f14p019 & 0.44\\
  4298 & f14p020 & 0.86\\
  4319 & f14p023 & 0.11\\
  4325 & f14p037 & 0.11\\
  4327 & f14p013 & 0.56\\
  4428 & f14p061 & 0.23\\
  4430 & f14p025 & 0.11\\
  4431 & f14p045 & 0.40\\
  4432 & f14p044 & 0.30\\
  4503 & f14p071 & 0.61\\
  4513 & f14p067 & 0.50\\
  4520 & f14p052 & 0.55\\
  4524 & f14p059 & 0.51\\
  4526 & f14p064 & 0.55\\
  4527 & f14p055 & 0.37\\
  4528 & f14p063 & 0.51\\
  4529 & f14p084 & 0.51\\
  4605 & f14p051 & 0.61\\
  4614 & f14p046 & 0.37\\
  4810 & f14p100 & 0.11\\
\hline\end{tabular}
\label{tab:literature_doubles}
\end{table}

The \citet{peng+04PN} PN and PN candidates include the \citet{hui+93b} sources 
(some of which have measured radial velocities from \citet{hui+95}). 
Hence it was decided to match the astrometry of the EMMI photometric 
catalogue using coordinates only from the Peng et al. catalogues. 
Additionally for the sources in common with the Hui et al. catalogues, 
5007\AA\ magnitude measurements from \citet{hui+93b} are reported.
Peng and collaborators already reported that PN 5420 and 5423 was the 
same PN listed twice in \citet{hui+95}. Two more pairs were also found 
within 1" in the \citet{hui+93b} list: 1219 and 4031 are separated by 
$0\farcs9$, and 5613 and 6112 by only $0\farcs4$. Additionally there are 
24 pairs within 1" among the PN candidates in Table 7 of 
\citet{peng+04PN}. All these pairs are listed in 
Table~\ref{tab:literature_doubles}. In principle it is possible that some 
of these are genuine separate objects, by chance aligned along the line of sight,  
but in a spectroscopic follow-up they would fall within the same fibre aperture (GIRAFFE MEDUSA
fibres have aperture diameter $1\farcs 2$). Therefore in the final catalogue all 
duplicate sources matching within 1$''$ are excluded.  The fact that all of the 24
close pairs from \citet{peng+04PN} are from the same field (f14) matched to
\citet{hui+93b} sources
points to a possible problem with the astrometry for that field. Furthermore it was 
subsequently confirmed that two PNe are the same source: the radial velocities of 
this pair (4430=f14p025 and 4614=f14p046), based on spectra taken with 
two different FLAMES configurations (one plate was configured taking the coordinates 
based on the ID reported in column one, the other based on ID and coordinates from 
column two in Table~\ref{tab:literature_doubles}), are identical to within the measurement 
errors.

Given a small difference in astrometry between the PN candidates and the confirmed PN from the 
Peng et al. catalogues (Figure~\ref{fig:radecEMMI_Peng}), we decided to match them 
independently to  the EMMI catalogues.
The initial match with the \citet{peng+04PN} catalogues identified 408 PN and 144 PN candidates 
within 1" from the EMMI PN candidates. The average offset between the two astrometric 
systems was found to be $0.36 \pm 0.19$ arcsec for the match with PN candidates, and slightly larger  
$0.38 \pm 0.19$ arcsec for the match with confirmed PN (Fig.~\ref{fig:radecEMMI_Peng}). 
To combine astrometry of the EMMI PN candidates with that of the catalogues from 
\citet{peng+04PN}, the coordinates of Peng et al. sources were shifted to 
match the EMMI astrometry. The shift applied to the PN candidates (Table 7 from Peng et al.) was  
$\Delta \mathrm{RA}= 0\farcs05$ and $\Delta \mathrm{DEC}=0\farcs2$, while 
$\Delta \mathrm{RA}=0\farcs21$ and $\Delta \mathrm{DEC}=0\farcs234$ was applied to PN (Table 6 
from Peng et al.).


Before making the final combined catalogue including all new EMMI PN candidates as well as 
confirmed PN and PN candidates from \citet{peng+04PN}, a selection of EMMI PN candidates 
was made based on the measured flux, removing sources brighter than the expected cut-off 
for the PN luminosity function measured by \citet{hui+93a}, i.e., 5007\AA\ magnitude $^{<}_{\sim}$ 23.0. 
To that purpose, the EMMI photometry was calibrated using the sources in common between 
the EMMI and \citet{hui+93b} catalogues.

\subsection{Photometry}

\begin{figure}
\centering
\resizebox{\hsize}{!}{
\includegraphics[angle=0]{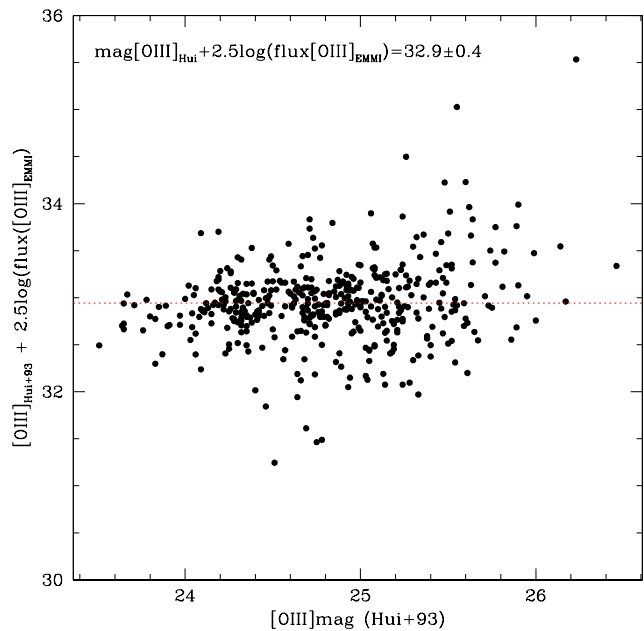}
}
\caption{Difference between the EMMI [O~III] - off-band filter aperture magnitude 
vs.\ the $m_{5007\AA}$ magnitudes from \citet{hui+93b} for PN candidates in common between
the two datasets.
}
\label{fig:EMMI_o3_GO_photo}
\end{figure}

Although the purpose of the EMMI imaging was to find PN candidates and not
to acquire accurate [O~III]5007\AA\ photometry, the observing conditions were
most of the time photometric and so this comparison was subsequently explored. 
For EMMI candidates in common with the photometric catalogue of 
\citet{hui+93b}, the aperture photometry flux measurements in the [O~III] 
image can be compared with the 5007\AA\ magnitudes. Fig.~\ref{fig:EMMI_o3_GO_photo}
shows the plot of the difference in the EMMI instrumental 5007\AA\ mag and the \citet{hui+93b} 
[O~III] mag for the 438 sources matched within 1" in both catalogues. 
The average difference is the zero point calibration for [O~III] EMMI photometry: 
\begin{equation}
\mathrm{ZPT} = \mathrm{[O~III]}_{Hui+93} + 2.5\log (Flux(\mathrm{[O~III]}_{EMMI})) = 32.9 \pm 0.4~\rm{mag}.
\end{equation}

The brightest PN in NGC~5128 have $\mathrm{[O~III]}_{Hui+93} \sim 23.5$~mag. Possibly a 
few PNe could be even brighter, but, for the purposes of pre-selection for the spectroscopic 
follow-up, a conservative limit of $m_{5007\AA}$ of 23.5 was adopted, corresponding to an 
observed $Flux(\mathrm{[O~III]}_{EMMI}) > 8000$ e$^{-}$. Therefore all sources with a total flux above
this limit (taking into account the scatter around the zero point) were rejected as being
too bright to be PNe. The faintest candidate has flux corresponding to [OIII] magnitude of 
$\sim 27$, which is plausible for PN luminosity in NGC~5128, and therefore no faint limit 
is imposed. After purging the bright sources, the EMMI catalogue contains 1803 PN candidates.
From this list, PN candidates were removed which were located within 1" of the sources for 
which spectra were obtained during the SV run, in order to form an input catalogue for the 
follow-up spectroscopy. On purpose some of the EMMI candidates for which Peng et al. obtained 
spectra were targeted also in the
open time observations, but none of the SV targets was re-observed during the FLAMES run. The 
photometric catalogue with all PN candidates from the new EMMI observations as well as from the 
literature are discussed in Section~\ref{sec:catalogs_imaging}.

\section{Spectroscopic data}
\label{sec:spectroscopy}

FLAMES, the Fibre Large Array Multi-Element Spectrograph is the 
multi-object, intermediate and high resolution spectrograph of the 
VLT with a 25 arcmin corrected field of view  \citep{pasquini+02}. 
In the MEDUSA mode it 
can observe up to 130 objects simultaneously through 1.2" fibres at 
a resolution R=6000 to 30000. It is thus ideally suited for kinematical 
studies of large samples of PN around nearby galaxies. 

FLAMES was employed to observe PN around NGC~5128 as part of the
FLAMES SV run in 2003, as well as within the regular observing
programme 073.B-0434 (GO) executed in 2004. The two 
observing runs are described here.

\subsection{FLAMES Science Verification Observations}

\begin{table*}
\caption[]{Log of SV VLT FLAMES GIRAFFE MEDUSA spectroscopy. The coordinates refer to the field centre and the airmass and DIMM seeing are taken from the headers. The 8$^{th}$ column lists the number of fibres allocated to PN candidates, and the last column the number of objects detected with emission lines, listing separately the number of globular clusters with detected emission lines.}
\centering 
\begin{tabular}{l l l l r l l l l} 
\hline\hline  
Field &  ~~~~~~~$\alpha$           & ~~~~~~~$\delta$ & Date & Exp & Airmass & DIMM &  \# Fibres on & \# Confirmed\\
      & (h ~~m ~s) & ~($^\circ$ ~~$'$ ~~$''$) &          & (s) &         & ($''$)   & PN cand. & PN+GCs w. emis. \\
\hline
Centre\_Field\_A1 & 13 25 28 & -43 00 49 & 2003-01-26 & 2180 & 1.106 & 0.83  & 105 & 97 + 4 GC \\
Centre\_Field\_A2 & 13 25 28 & -43 00 49 & 2003-01-27 & 2180 & 1.453 & 1.06  & 105 & 98 + 4 GC \\
Centre\_Field\_B1 & 13 25 28 & -43 00 49 & 2003-01-27 & 2180 & 1.266 & 0.77  & 111 & 101 + 2 GC  \\
Centre\_Field\_B2 & 13 25 28 & -43 00 49 & 2003-01-27 & 2180 & 1.164 & 1.35  & 111 & 104 + 2 GC  \\
Centre\_Field\_C1 & 13 25 28 & -43 00 49 & 2003-01-28 & 2180 & 1.141 & 1.03  & 110 & 105 + 1 GC \\
Centre\_Field\_C2 & 13 25 28 & -43 00 49 & 2003-01-30 & 2180 & 1.090 & 1.19  & 111 & 106 + 1 GC \\
Centre\_Field\_D1 & 13 25 28 & -43 00 49 & 2003-01-29 & 2180 & 1.245 & 1.18  & 109 & 102 + 1 GC \\
Centre\_Field\_D2 & 13 25 28 & -43 00 49 & 2003-01-29 & 4200 & 1.127 & 1.64  & 110 &  104 + 1 GC \\
Centre\_Field\_E1 & 13 25 28 & -43 00 49 & 2003-01-30 & 2180 & 1.378 & 1.09  & 101 &96 + 4 GC  \\
SW\_Field         & 13 24 45 & -43 09 53 & 2003-02-02 & 2180 & 1.079 & 0.61  &  76 & 70 + 1 GC \\
\hline
\end{tabular}
\label{tab:SVObs}
\end{table*}

During the FLAMES SV campaign, FLAMES in UVES+MEDUSA 
combined mode was used to observe known PN and globular clusters (GCs) in NGC~5128: 
this combined mode allowed parallel observation of a single setup of 130 targets 
with MEDUSA fibres ($1\farcs 2$ fibre aperture) and up to 8 additional targets at 
very high resolution (R=47000) through 8 $1\farcs 0$ fibres, which
are fed to the red arm of UVES, recording spectra from 476-683 nm. 
The 130 MEDUSA fibres are fed to the GIRAFFE 
spectrograph that was used with the low resolution L3 (R=7500) grating, 
covering the spectral range from 450-508 nm. Simultaneous ThAr calibration 
lamps were switched-off during the science exposures.

Between 25 January  and 2 February 2003, six different FLAMES MEDUSA set-ups 
were observed in 2 different fields, in the centre and SW part of the galaxy 
(programme: 60.A-9140).  The central field, due to its large source density, 
had 5 different set-ups but the SW field only one. Table~\ref{tab:SVObs} lists 
the field coordinates, observing date, exposure time, airmass  and DIMM seeing
at the start of the exposure. The last two columns list the number of fibres 
allocated to PN candidates and the number of detected PN.

Due to lack of time one additional 
planned setup in the NE of the galaxy could not be taken. In total,
during the FLAMES SV run, GIRAFFE spectra were taken for 612 PN candidates, 
confirming 527 individual PN\footnote{Originally 614 PN were observed, confirming 529, 
but two pairs (5613=6112 and 1219=4031 where IDs come from \citet{hui+93b}) were later 
identified as being the same source.}. Of these 166 are newly confirmed PN, as 
they did not have radial velocities in the literature. Additionally 48 globular 
cluster spectra were collected within these set-ups.  For each setup two exposures 
of 2180 sec were taken, except for one central field and the SW field which both
received only one such exposure. Thus the total science time was $10 \times 45$~min=7.5~h. 
In addition, six bright GCs were observed with UVES 580 nm central wavelength 
setting in all the exposures of the central field. 

PN emission was detected in 11 globulars \citep{rejkuba+walsh06} and these sources are
also discussed below. The spectra of other GCs obtained with MEDUSA and UVES fibres 
are not discussed further here.

PNe were selected from the \citet{hui+93b} catalogue and the GCs are from 
various published sources. The observed PNe cover a range in $m_{5007\AA}$ from 
23.5 to 26.8 mag. The imaging material, on which the astrometric catalogue
for the SV observations is based, was compiled from an MPG/ESO 2.2m WFI 
survey of NGC 5128 (proposal 
67.B-0111(A), epoch 2001). A mosaic of 6 WFI pointings observed with an 
[O~III]5007\AA, filter (OIII/2, central wavelength 502.612 and width 2.818nm) 
had been obtained. Whilst most of the catalogued PN are detectable on this 
mosaic, the PN catalogue of \citet{hui+93b} was transferred to the 
epoch of this image by $\alpha, \delta$ shift. The image was matched against 
GSC and secondary reference stars and the target coordinate information 
referred to the [O~III] image.

All the VLT SV data are public and the FLAMES data are available from 
{\tt http://www.eso.org/science/vltsv/flamessv/}.

\subsection{FLAMES Period 73 observing run}

\begin{table*}
\caption[]{Log of VLT FLAMES GIRAFFE MEDUSA spectroscopy taken within the service mode 
observations 073.B-0434(B).}
\centering
\begin{tabular}{l l l l r c l l l} 
\hline\hline 
Field &  ~~~~~~~$\alpha$           & ~~~~~~~$\delta$ & Date & Exp & Airmass & DIMM & \# Fibres on & \# Confirmed\\
      & (h ~~m ~s) & ~($^\circ$ ~~$'$ ~~$''$) &          & (s) &         & ($''$) & PN cand. & PN  \\
\hline
 FIELD1  & 13 26 20 & -42 39 20 & 2004-06-13 & 2x2595 & 1.054 & 0.88   &   95 & 50 \\
 FIELD2  & 13 26 30 & -42 50 20 & 2004-06-16 & 2x2595 & 1.133 & 0.88   &  113 & 60 \\
 FIELD3  & 13 24 22 & -43 15 34 & 2004-06-17 & 2x2595 & 1.117 & 0.71   &   66 & 50 \\
 FIELD4  & 13 24 40 & -43 01 41 & 2004-06-17 & 2x2595 & 1.067 & 0.76   &  114 & 72 \\
 FIELD5  & 13 26 15 & -42 49 59 & 2004-06-18 & 2x2595 & 1.083 & 0.66   &  106 & 61 \\
 FIELD6  & 13 24 25 & -42 51 11 & 2004-06-18 & 2x2595 & 1.265 & 0.77   &   60 & 43 \\
 FIELD7\_1  & 13 26 07 & -42 54 19 & 2004-06-18 & 2595 & 1.077 & 0.51  &   98 & 60 \\
 FIELD7\_2  & 13 26 07 & -42 54 19 & 2004-06-28 & 2595 & 1.206 & 1.02  &   98 & 52 \\
 FIELD8  & 13 25 27 & -43 02 55 & 2004-06-28 & 2x2595 & 1.089 & 0.89   &  115 & 75 \\
 FIELD9  & 13 25 27 & -43 02 55 & 2004-06-28 & 2x2595 & 1.502 & 1.20   &  114 & 57 \\
 FIELD10 & 13 25 51 & -43 03 26 & 2004-07-11 & 2x2595 & 1.082 & 0.75   &  115 & 64 \\
\hline
\end{tabular}
\label{tab:GOObs}
\end{table*}

For the FLAMES observing run 073.B-0434(B), the same MEDUSA GIRAFFE 
setup L3 (R=7500), covering the spectral range from 450-508 nm, was  
adopted as for the SV data collection.  However this time the 
simultaneous ThAr calibration lamp was switched on in order to achieve as 
high as possible wavelength calibration accuracy, and the UVES fibres were not used. 
All targets were PN or PN candidates either from our own EMMI imaging, 
or from the \citet{peng+04PN} catalogues (their Table 6 and 7), which include sources
from the \citet{hui+93b}. 

Observations were taken in service mode. 
Table~\ref{tab:GOObs} lists the FLAMES field coordinates, observing date, 
exposure time, airmass and DIMM seeing at the start of the exposure. The last 
two columns list the number of fibres allocated to PN candidates and the 
number of detected PN, respectively. In total 996 sources were targetted and 580 
confirmed as PN members in NGC~5128. Of these, 297 
were newly confirmed PN with single velocity components. 
No targets in common were observed during the two spectroscopic runs. 

\begin{figure}
\centering
\resizebox{\hsize}{!}{
\includegraphics[angle=0]{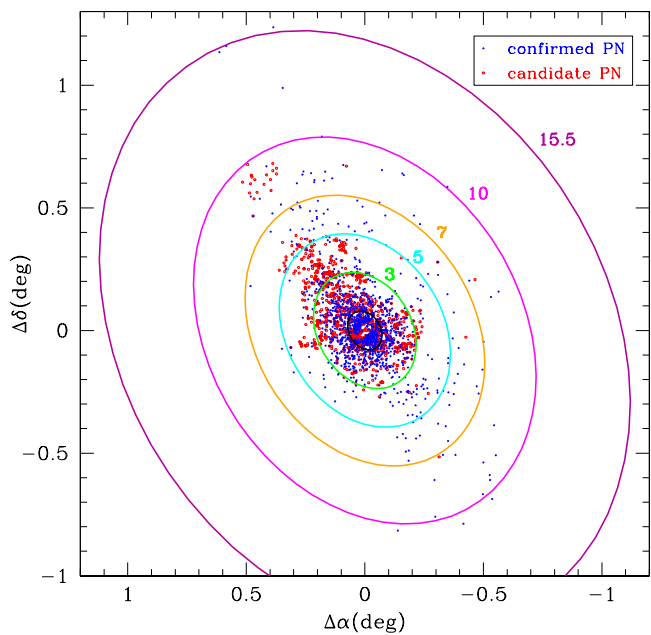}
}
\caption{
Distribution of spectroscopically confirmed PN (blue filled dots) and PN candidates 
(open red circles) from EMMI imaging and the literature. The origin of the coordinate system 
is centred on NGC~5128, the elliptical isophotes have axis ratio b/a=0.77 with position 
angle 35$^\circ$ and the innermost ellipse is drawn at 1 effective radius ($1R_{eff}$ = 305$''$
\citep{dufour+79}, equivalent to 5.62 kpc); the size of the other ellipses is 3, 5, 
7, 10 and 15.5 effective radii as indicated. 
}
\label{fig:Targets}
\end{figure}

Figure~\ref{fig:Targets} displays the spatial distribution of all the observed 
PN (red filled dots) and PN candidates (open black circles), with the origin of the coordinate system
centred on NGC~5128 (taken as the NED coordinate: $13^{h} 25^{m} 27\farcs 6$,  
$-43^{\circ} 01' 09''$ [J2000]). 

\subsection{GIRAFFE reductions}
\label{sec:rv}

FLAMES GIRAFFE data reduction was performed using the girBLDRS pipeline \citep{Blecha+00}. For 
bias subtraction, a master bias frame was created using the {\it biasMast} task. Given that 
the master bias provided with the pipeline package was essentially indistinguishable from the 
master bias created using our data, the former was adopted. As a first preparation step for 
the extraction of science spectra, the {\it locMast} task was run, which takes as input the 
temporally closest flat-field calibration files and produces the localization file - a 
description of the trace of the spectrum of each fibre.  Then the master wavelength 
calibration file was created with {\it wcalMast}. Finally the extracted, wavelength 
calibrated science spectra were produced with the {\it extract} pipeline task. This task 
subtracts a 2D scattered light polynomial model, identifies the spectra and extracts them. 
Furthermore it applies a flat-field correction and re-bins the spectra to  a linear wavelength 
scale. Finally the wavelength solution is adjusted through cross-correlation with the five 
simultaneous ThAr calibration fibres present in the science images. The latter step was only 
applicable for the open time run 073.B-0434(B), since the simultaneous calibration lamp was 
not used during the Science Verification run. No sky subtraction was applied due to very low 
sky background in virtually all spectra and the lack of any strong telluric features at the 
wavelengths of the PN emission lines. A few experiments with sky subtraction, where the mean 
sky spectrum was created from a combination of 10-20 spectra allocated to empty sky positions, 
showed that, as far as the velocity measurements were concerned, the results were unchanged. 

The extraction algorithm applied was a direct extraction (SUMM), because optimal extraction 
was found to remove most of the PN emission lines due to their similarity with cosmic-rays. 
In consequence the final extracted and wavelength calibrated spectra still contained many 
cosmic-rays traces but these were
detected during the line fitting process either by eye, as narrower than the PN emission
lines, or from their anomalous velocity.

\subsection{Line fitting}
The PN emission lines were fitted by Gaussians in order to determine the (single) 
recession velocity of each detected PN. Line fitting was done independently using
two methods. Manual fitting was performed with the {\it splot} task in IRAF for as 
many available lines that could be distinguished by eye. The error array provided by
girBLDRS pipeline for each extracted spectrum was used to estimate the 
appropriate scalar to transform the input signal to pixel sigma as required by 
{\it splot}. Fitting of a Gaussian was performed with 100 Monte Carlo trials
to produce error estimates on the line wavelength, width and flux. In all
cases the line fitting was done independently on two exposures for each field.
This cautious approach was deemed to be vindicated by the occasional detection of 
a weaker line in the better exposed spectrum of a pair and the not-uncommon
presence of cosmic rays within the emission line extent. The final velocity 
measurement of each PN is based on the weighted mean (weighting by 
the relative line strength) of all the
detected lines (thus at least [O~III]5007\AA\ and generally with 4959\AA, but
also in some cases with H$\beta$) from the two independent fits. The velocity 
error was propagated from this combination.

An independent fitting with an automatic code was also performed as a check.
In this case the pairs of spectra were combined using {\it IRAF.imcombine} to
remove cosmic rays and then lines were autonomously detected above the flat
continuum based on a detection threshold. The code was applied without
an estimate of the radial velocity or likely line ratios and detected 
less than 50\% of the lines found manually. However in the cases where
the identical lines were fitted, the agreement with the {\it splot} results
for the fitting of Gaussian emission lines was found to be excellent and within 
the error bars. The {\it splot} manual fit results are quoted here.

\section{NGC~5128 Planetary Nebulae Catalogues}
\label{sec:catalogs}

\subsection{Catalogue of planetary nebula candidates}
\label{sec:catalogs_imaging}

\begin{table*}
\caption{PN candidates from EMMI imaging run. Column 5 and 6 are the ID and 5007\AA\, magnitude
 from \citet{hui+93b} and the last column is the ID from \citet{peng+04PN} for sources found in 
 common within 1'' matching radius. Here only a few lines are shown as an example. The complete 
 catalogue is available in electronic form.}
\begin{tabular}{lrrllll}
\hline\hline
ID(EMMI) & $\alpha$~~~~~~ & $\delta$~~~~~~ & m(5007\AA)$_{\mathrm{EMMI}}$ & 
ID(H+93) & m(5007\AA)$_{\mathrm{H+93}}$ & ID(P+04) \\
         &  (~h ~~m ~~s~) & (~$^\circ$ ~~~$\arcmin$ ~~~$\arcsec$~) & & & & \\ 
\hline
  EMMI\_956  & 13 25 39.71 & -43 00 27.9 & 23.42  &   253  &  25.55  &    \\
  EMMI\_1176 & 13 26 06.10 & -43 09 06.9 & 24.97  &  5506  &  25.00  &    \\
  EMMI\_1696 & 13 25 09.71 & -43 03 41.3 & 24.91  &  4243  &  24.81  &    \\  
  EMMI\_952  & 13 25 37.52 & -43 00 16.3 & 25.48  &        &         &     \\      
  EMMI\_954  & 13 25 36.70 & -43 00 21.4 & 24.7   &        &         &      \\        
  EMMI\_1708 & 13 25 04.14 & -43 04 23.5 & 24.71  &        &         & f14p034 \\
  EMMI\_1729 & 13 25 04.91 & -43 05 31.2 & 24.29  &        &         & f14p026 \\
\hline
\end{tabular}
\label{tab:EMMI_PNcand}
\end{table*}

The combined list of all EMMI PN candidates, culled of duplicate sources 
within 1'' and sources which are too bright to be PNe 
($m(5007\AA)_{\mathrm{EMMI}}$ $^{<}_{\sim} ~23.0$), contains 1803 entries. 
Of these 718 have radial velocities measurements from FLAMES spectroscopy 
and 410 have radial velocities reported by \citet[][Table 6]{peng+04PN}. 
Since FLAMES spectra were also taken of PN from \citet{peng+04PN}, the total 
number of new PN candidates with no RV measurements is 1060.
Table~\ref{tab:EMMI_PNcand} shows the first 5 lines of the EMMI catalogue of 
1060 PN candidates, including the IDs and magnitudes from \citet[][H+93]{hui+93b} 
and IDs \citet[][P+04]{peng+04PN} catalogues. Magnitudes from EMMI imaging are 
based on total flux measured in [O~III] filter calibrated using the sources 
found in common with \citet{hui+93b}; see Fig.~\ref{fig:EMMI_o3_GO_photo}. 
The complete table is available from CDS.

\begin{table*}
\caption{PN candidates from literature \citep[][Table 7]{peng+04PN} that have no 
radial velocity measurements to date (58 sources). 
The RA and DEC are shifted to match the astrometric system of Table~\ref{tab:EMMI_PNcand}. 
ID and 5007\AA, magnitude from \citet{hui+93b}  are also reported, where available. Here 
only a few lines are shown as an example. The complete catalogue is available in electronic form.}
\begin{tabular}{lrrll}
\hline\hline
ID(P+04) & $\alpha$~~~~~~ & $\delta$~~~~~~ & ID(H+93) & m(5007\AA)$_{\mathrm{H+93}}$  \\
         &  (~h ~~m ~~s~) & (~$^\circ$ ~~~$\arcmin$ ~~~$\arcsec$~) & &  \\
\hline
  1218 &  13 26 03.90 & -42 53 20.8 & 1218 & 25.62 \\
  1603 &  13 26 26.99 & -42 45 02.2 & 1603 & 26.36 \\
  1801 &  13 25 39.85 & -42 41 04.6 & 1801 & 25.28 \\
  2201 &  13 23 59.64 & -43 16 31.2 & 2201 & 24.61 \\
 f04p5 &  13 23 44.36 & -43 32 02.9 &      &       \\
\hline
\end{tabular}
\label{tab:PNcand_literature}
\end{table*}

Table~\ref{tab:PNcand_literature} reports the first 5 lines for the remaining 58 PN 
candidates from the literature. The complete table is available from CDS. These are
PN candidates from table 7 of \citet{peng+04PN}  that had no spectroscopic confirmation yet, and that were not 
covered by the EMMI imaging. 
Their coordinates are shifted to match the EMMI astrometric system by adding 0.05 arcsec to RA 
and 0.2 arcsec to DEC. Additionally the photometry of PN candidates in common with 
\citet{hui+93b} was matched in order to report consistent 5007\AA\, magnitudes. 

Summing the 1060 PN candidates from NTT EMMI imaging with the 58 candidates from the 
literature, there are a total of 1118 PN candidates in NGC~5128 awaiting spectroscopic 
confirmation. Their spatial distribution is plotted with open red circles in Figure~\ref{fig:Targets}. 

\subsection{Catalogue of spectroscopically confirmed planetary nebulae}
\label{sec:catalogs_spectro}


\begin{table*}
\caption{Radial velocity, radial velocity error and 5007\AA\ line FWHM measurements for 1107 
single velocity component PNe observed with FLAMES within the SV and GO runs. In addition identifiers 
from EMMI, \citet[][H+03]{hui+93b} and \citet[][P+04]{peng+04PN} catalogues, 5007\AA\ magnitudes from 
EMMI and \citet{hui+93b} are listed, as well as radial velocities from \citet{peng+04PN}, for those PN in 
common between the matched catalogues. Here only few lines are shown as example.
Note that the coordinates of the PN observed in the SV run are offset by $\Delta \alpha = -0.225''$ 
and $\Delta \delta=-0.093''$ with respect to the EMMI astronometry that was used 
for the GO run. The complete catalogue is available in electronic form.}
\begin{tabular}{lccccclclclc}
\hline\hline
Run & $\alpha$ & $\delta$ & RV & $\sigma_{RV}$   & FWHM  &ID&m(5007\AA)& ID &m(5007\AA)& ID & RV (P+04) \\ 
    & (~h ~m ~s~) & (~$^\circ$ ~~$\arcmin$ ~~$\arcsec$~) & (kms$^{-1}$) & (kms$^{-1}$) &  5007\AA & EMMI & (EMMI) & (H+93) & (H+93)& (P+04) & (kms$^{-1}$) \\
\hline
  GO & 13 26 18.89 & -42 31 34.0 & 590.0 & 0.8 & 37.2 &             &       &       &       & f22p22   & 575 \\    
  GO & 13 26 02.41 & -42 39 05.8 & 532.5 & 0.9 & 44.2 & 2483  & 25.31 &       &       & f23p08   & 528 \\      
  GO & 13 25 06.68 & -42 58 16.2 & 707.0 & 1.3 & 37.1 &       &       & 4430  & 25.27 & 4430=    &     \\
     &             &             &       &     &      &       &       &       &       & f14p025  &     \\  
  GO & 13 26 16.32 & -42 38 21.4 & 623.0 & 3.7 & 44.7 & 2460  & 25.94 &       &       &          &     \\  
  SV & 13 24 50.40 & -43 03 22.2 & 452.5 & 4.3 & 73.5 &             &       &       &       &          &     \\
  SV & 13 23 46.78 & -43 16 23.2 & 585.2 & 1.9 & 61.0 &             &       & 2801  & 25.12 & f10p8    & 567 \\ 
  SV & 13 24 21.77 & -42 59 28.6 & 511.8 & 1.1 & 48.6 & 585   & 24.74 & 4805  & 24.70 &          & 517 \\ 
  SV & 13 25 02.91 & -42 59 38.1 & 749.2 & 0.8 & 53.0 & 340   & 24.50 & 4415  & 24.65 & ctr415   & 657 \\         
\hline         
\end{tabular}
\label{tab:PN_ALL_vels}
\end{table*}

Table~\ref{tab:PN_ALL_vels} lists the radial velocity, radial velocity errors and 5007\AA\ line 
FWHM measurements for 1107 single velocity component PNe observed with FLAMES within the SV and 
Period 73 FLAMES runs. In addition to these PNe, there are 9 targets, displaying double or even 
more complex velocity profiles in their emission lines; these are discussed in more detail in Section~\ref{sec:doublePN}. Also (single velocity component) emission lines were detected in 
9 globulars; these are listed in Tab.~\ref{tab:GCPN}. On the basis of these emission line 
measurements, a total of 486 PNe are newly confirmed, extending the work of \citet{hui+95} and \citet{peng+04PN}.

For all PN except five, the listed values are based on the averaged measurements from 
two exposures made with the identical configurations, as described above. The five PN have values averaged 
from 4 exposures based on 2 independent plate configurations. This is due to the fact that these 5 PN were observed
each as two originally different sources (the sources had different input ID's and their coordinates differed slightly, 
but still within 1" on the sky and hence within the FLAMES MEDUSA fibre aperture). 
The four pairs  came from the PN candidates from the literature: 
PN 4430  = f14p025, and  4614 = f14p046 (these were taken as 4 independent PN candidates from the 
Table 7 of  \citet{peng+04PN}), and the following 4 PNs (2 pairs) were taken from \citet{hui+93b} 5613=6112 
and 1219=4031. These were already found to be likely double 
detections in Section~\ref{sec:astrometry} (see in particular Table~\ref{tab:literature_doubles}),
and velocity measurements confirm this. One additional pair of PNe within 1" (EMMI\_63 and EMMI\_683) comes 
from the EMMI imaging, due to the fact that the input list for the spectroscopic follow-up had only been 
matched to $0\farcs 5$.\footnote{Here for the presentation of the final PN candidate catalogue 
(Tab.~\ref{tab:EMMI_PNcand}), the internal match was made to 1"; see Sec.~\ref{sec:astrometry}.}. 
Hence we only list one source (EMMI\_63), but given that this PN was observed on two configurations, 
again its final velocity, error and flux are the averages from the 4 exposures based on two independent configurations.  

Observed wavelengths were converted to heliocentric radial velocity applying the heliocentric correction
appropriate for each observation and standard rest wavelengths. Table~\ref{tab:PN_ALL_vels}  
lists the run (SV or Period 73 open time, marked as 'GO') identifying in which run 
each observation was taken, the RA and Dec taken from the FLAMES configuration files,
the mean radial velocity and error, the FWHM of the 5007\AA\ line as well as 
IDs from EMMI, \citet{hui+93b} and \citet{peng+04PN} catalogues. 
There is a small systematic offset between the SV coordinates and the EMMI astrometry.
In order to place the SV coordinates on the same positional system as the EMMI 
coordinates, offsets of $\Delta RA = -0.225''$ and $\Delta DEC = -0.093''$ should be 
applied to the SV coordinates. Additionally the magnitudes from EMMI and 
from \citet{hui+93b}, and the velocities from \citet[][Table 6]{peng+04PN} for 
those PN in common, are listed. Of the 1107 PN for which radial velocities are listed 
in Table~\ref{tab:PN_ALL_vels}, 166 are newly confirmed PN from the SV run and 297 
from the later open time run, increasing the number of confirmed PN by 463 in NGC~5128.

\begin{figure}
\centering
\resizebox{\hsize}{!}{\includegraphics[angle=-90]{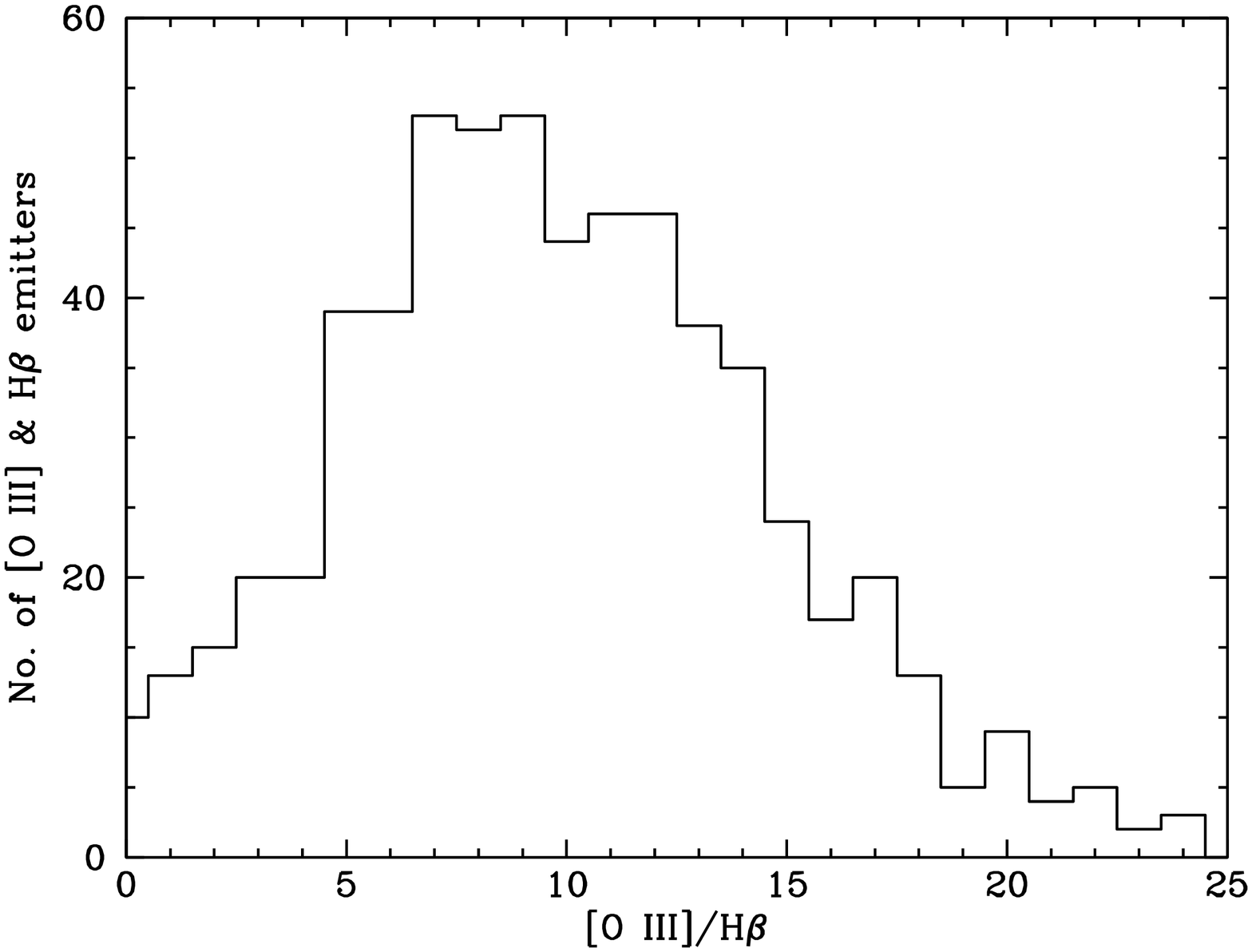}}
\caption{Histogram of the distribution of the [O~III]/H$\beta$ ratio for the
629 PN for which both lines were detected at $^{>}_{\sim} 2 \sigma$. An additional
9 measurements lie beyond the upper limit of this plot.
}
\label{fig:Hist_o3hb}
\end{figure}

Table~\ref{tab:PN_ALL_fluxes} lists the counts in the 5007\AA\ line and the 
[O~III]/H$\beta$ flux ratio (in the cases where H$\beta$ was detected), for all the 
single velocity component PNe. The IDs have been abbreviated in this case - only the ID of the 
source that was taken as input for the spectroscopy is shown (EMMI or 
\citet[][Table 6 or 7]{peng+04PN} for open time observations
and \citet{hui+93b} for the SV observations). In total 629 PN (57\%) had H$\beta$ 
detected. A histogram of the [O~III]/H$\beta$ ratio is shown in Fig.~\ref{fig:Hist_o3hb};
the mean value is 10.2 and standard deviation of the mean 5.5. 
In addition a small number (34) of PNe had the He~II 4686\AA\ line detected
(at $> 2 \sigma$) and the He~II/H$\beta$ ratio is listed in Table~\ref{tab:PN_ALL_fluxes}: 
the mean He II/H$\beta$ ratio was 0.48.  

\begin{table*}
\caption{Relative flux of the [O~III] 5007\AA\ line, the [O~III]/H$\beta$ and
the He~II/H$\beta$ line ratio measurements (where H$\beta$ and He~II lines detected)
for all PN observed with FLAMES 
within the SV and GO runs (c.f., Tab.~\ref{tab:PN_ALL_vels}). The name contains a 
pre-fix that indicates whether the ID is taken 
from EMMI imaging, or from pre-existing catalogues (Hui93 for SV run and P04 for GO run 
observations). All alternative IDs may be found in Table~\ref{tab:PN_ALL_vels}. Here only a few 
lines are shown as example. The complete catalogue is available in electronic form.}
\begin{tabular}{lcccrr}
\hline\hline
 Name	& $\alpha$   & ~~~$\delta$ & Rel. flux &    [OIII]/H$\beta$ & He~II/H$\beta$\\
        & (~h ~~m ~~s~) & (~$^\circ$ ~~~$\arcmin$ ~~~$\arcsec$~) &  5007\AA &    &                               \\
\hline           
P04\_f22p18 &   13 26 25.53 & -42 38 21.5 &   52.3   &  12.63 & 0.51 \\
P04\_4430   &	13 25 06.68 & -42 58 16.2 &   23.3   &   6.72 &   \\
Hui93\_5613 &	13 25 41.83 & -43 07 24.9 &   26.3   &  12.17 &   \\   
Hui93\_1219 &	13 25 56.84 & -42 53 50.5 &   13.9   &        &   \\
P04\_f22p22 &	13 26 18.89 & -42 31 34.0 &   23.5   &   4.39 &  0.38 \\   
EMMI\_2460 &	13 26 16.32 & -42 38 21.4 &   10.6   &        &   \\
EMMI\_2483 &	13 26 02.41 & -42 39 05.8 &   26.8   &   8.37 &   \\   
\hline
\end{tabular}
\label{tab:PN_ALL_fluxes}
\end{table*}

\begin{figure}
\centering
\resizebox{\hsize}{!}{\includegraphics[angle=-90]{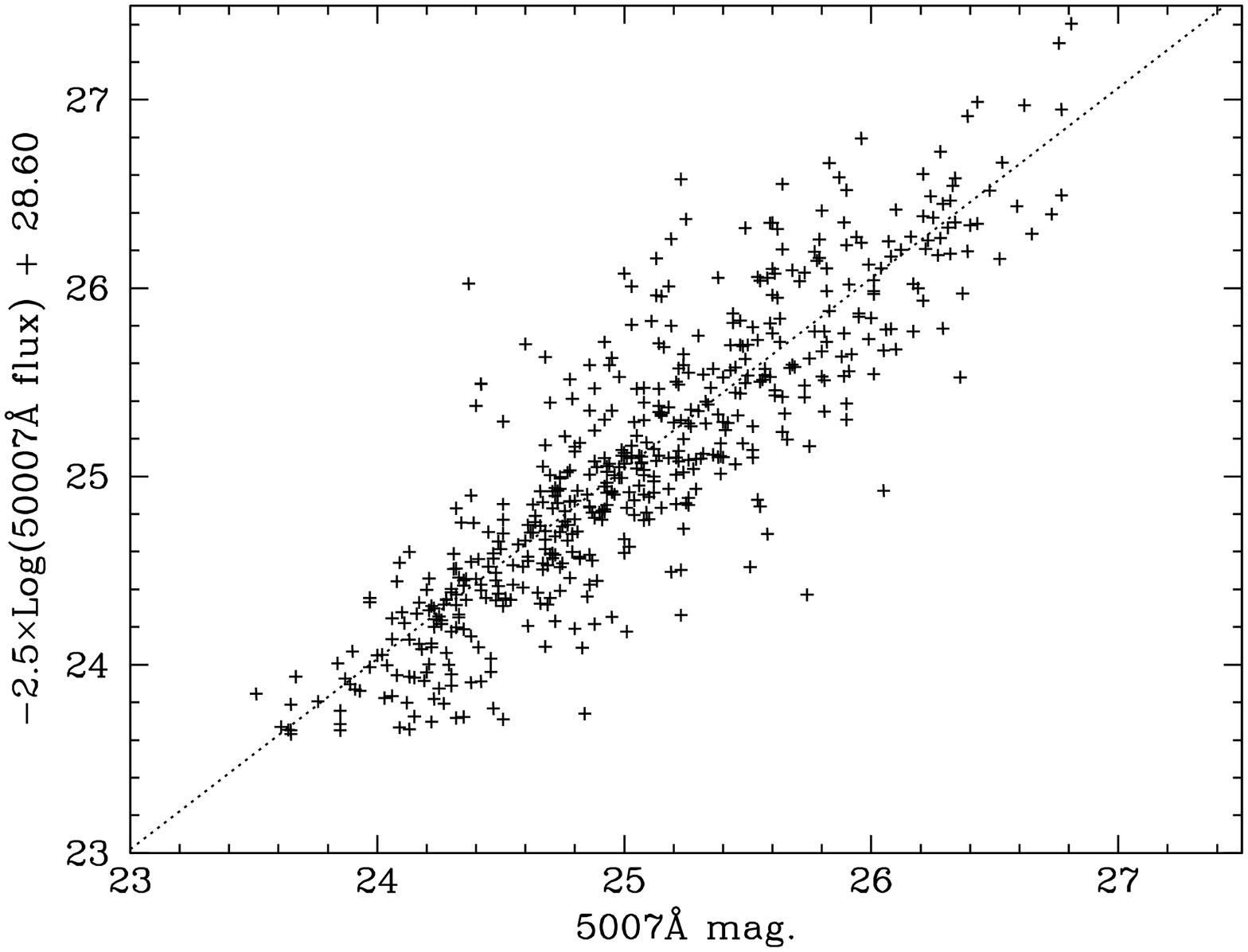}}
\caption{Correlation of the log of the fitted 5007\AA\ line flux against m$_{5007A}$ 
for those PN from \citet{hui+93b} which were observed with FLAMES in the SV 
programme. The least squares linear fit is shown.  
}
\label{fig:Corr_FluxMag}
\end{figure}

The strength of the 5007\AA\ line is a useful relative flux indicator among
the observed PN, especially as conditions were generally photometric or clear
during the FLAMES observations. For 700 PN observed with FLAMES that 
have an $m_{5007\AA}$ from \citet{hui+93b}, a plot of $ log (Flux(5007\AA)$ $v.$ 5007\AA\ mag.
shows an excellent correlation, with however some large outliers (Fig. 
\ref{fig:Corr_FluxMag}). A linear fit shows a slope of 1.011$\pm$0.025,  the 
zero point is $28.59 \pm .42$, and the rms scatter around the fit is 0.35. 
This shows that a similar level of photometric calibration
could be obtained from FLAMES spectra to provide m$_{5007A}$, as with the  
photometry of the PN detected with EMMI. 

\begin{figure}
\centering
\resizebox{\hsize}{!}{\includegraphics[angle=-90]{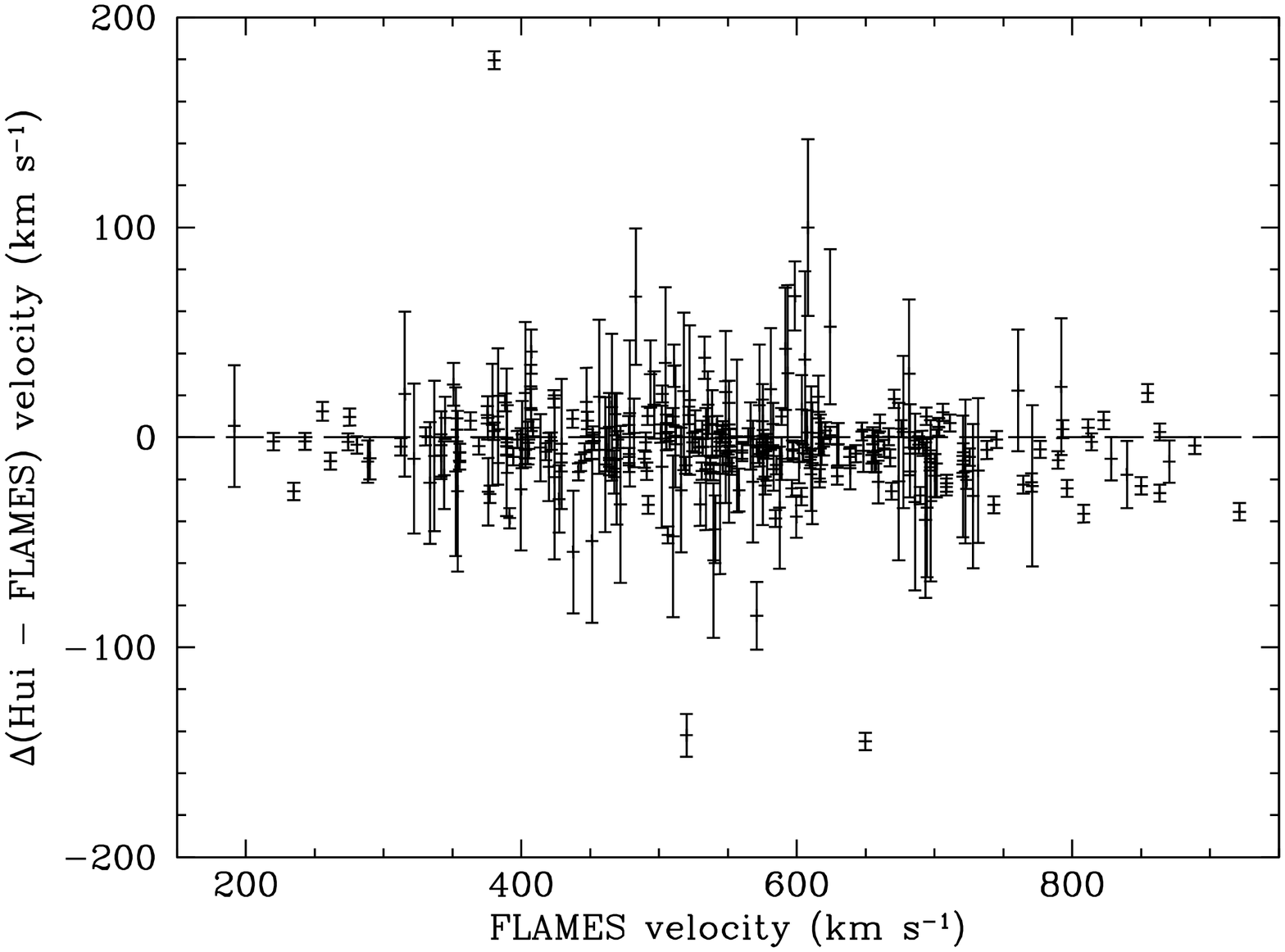}}
\resizebox{\hsize}{!}{\includegraphics[angle=-90]{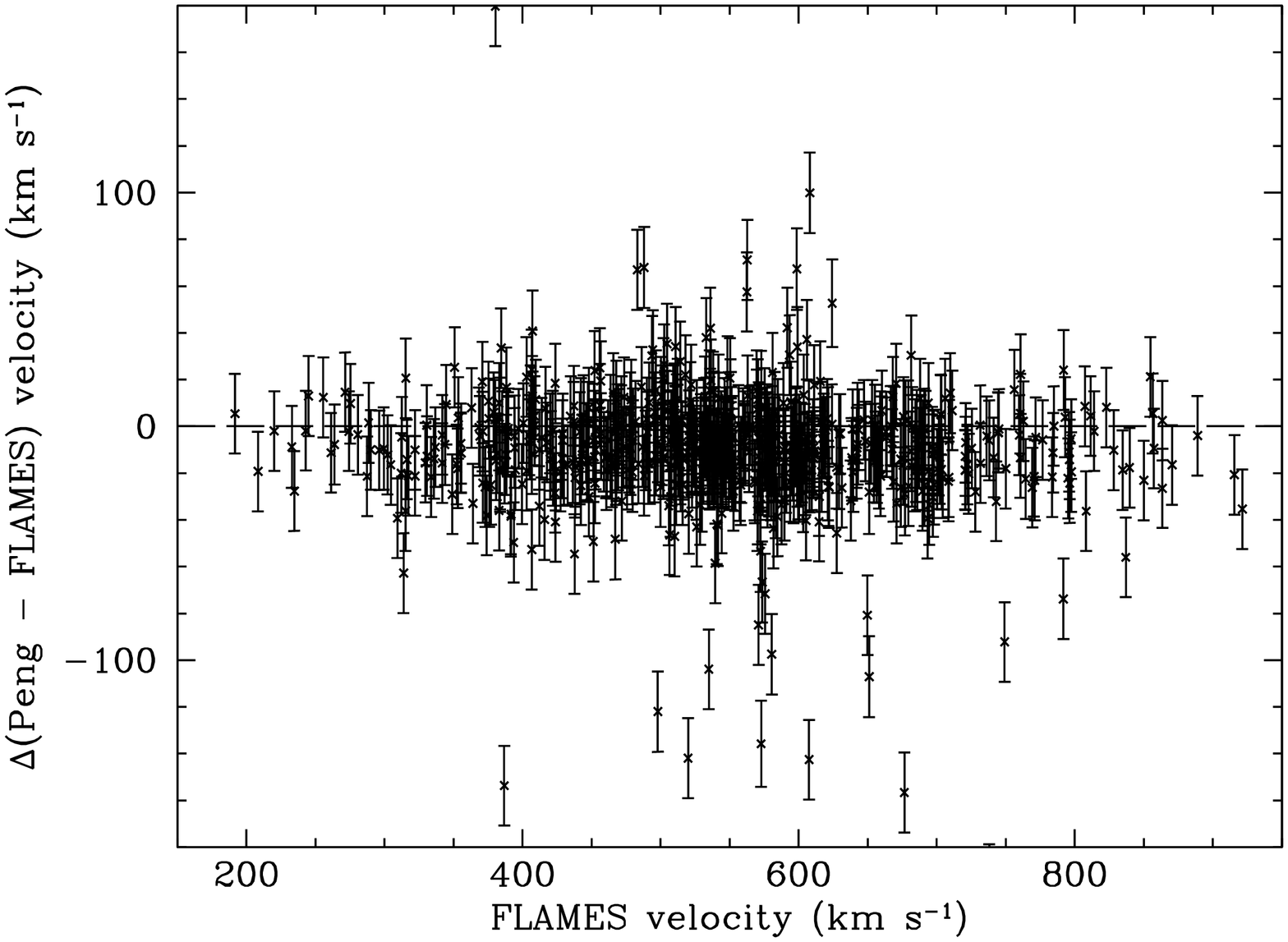}}
\caption{
Upper panel: Offset between the radial velocity of the PNe in common between the 
\citet{hui+95} and the FLAMES observations. The mean 
offset (Hui - FLAMES) and standard error is $-3.7 \pm 31.7$ kms$^{-1}$ (291 PN in common). The
error bars are the quadratic sum of the FLAMES and Hui errors. \newline
Lower panel: Offset between the radial velocity of the PNe in common between the 
\citet{peng+04PN} observations and the FLAMES spectra is shown. The mean 
offset (Peng - FLAMES) and standard error is $-9.9 \pm 41.5$ kms$^{-1}$ for 645 spectra
in common between Peng, SV and P73 FLAMES programmes. After removing 36 matches
with the most deviant offsets ($| \Delta | >50$ kms$^{-1}$), the mean offset (Peng - FLAMES) and in 
particular its standard deviation are significantly lower: $-7.9 \pm 15.5$ kms$^{-1}$.
The error bars are the quadratic sum of the FLAMES errors and an assumed 17.0 kms$^{-1}$ error for
measurements given in Table 6 of \citet{peng+04PN}.}
\label{fig:FLAMES_othervels}
\end{figure}

The comparison of the radial velocity measurements from the FLAMES spectra 
with those from \citet{hui+95} for the SV data, and the \citet{peng+04PN}
data in common with both SV and Period 73 open time FLAMES velocities are shown in Fig. \ref{fig:FLAMES_othervels}.
For the \citet{hui+95}   dataset, the mean offset is very low ($-3.7$ kms$^{-1}$,
in the sense Hui - FLAMES velocity) but the spread is considerable with some large
outliers. For the \citet{peng+04PN} data (after removing 37 most deviant points with 
differences $>50$ kms$^{-1}$), the spread is lower but the offset 
larger in magnitude ($-7.9$ kms$^{-1}$, in the sense Peng - FLAMES velocity). 
Since the FLAMES spectra generally have
very low errors on account of the strong signal and moderate spectral resolution 
($\sim$40 kms$^{-1}$), and there is excellent consistency between velocities from
[O~II]4959\AA\ and H$\beta$, where detected, and [O~III]5007\AA, the FLAMES velocities
are superior to those from the lower resolution observations of Hui et al.
and Peng et al.  

\begin{table*}
\caption{132 PNe with radial velocity measurements from \citet[][Table 6]{peng+04PN} that 
were not observed with FLAMES. The velocities listed are taken directly from the \citet{peng+04PN} 
catalogue and should be shifted by +7.9 kms$^{-1}$ when combining with the FLAMES velocities in 
Table~\ref{tab:PN_ALL_vels} in order to produce a single coherent dataset. The RA and DEC are shifted 
to match the astrometric system of Table~\ref{tab:EMMI_PNcand}. Here only a few lines are shown as an 
example. The complete catalogue is available in electronic form.}
\begin{tabular}{lccclclc}
\hline\hline
 ID	      &	~~~$\alpha$	      &  ~~~$\delta$  &     RV(P+04)   &    ID          & m(5007\AA) & ID  & m(5007\AA) \\
 (P+04)  & (~h ~~m ~~s~) & (~$^\circ$ ~~~$\arcmin$ ~~~$\arcsec$~) & (km~s$^{-1}$)  &   (H+93)  & (H+93) &  (EMMI) & (EMMI)   \\
\hline           
  ctr132  & 13 25 11.20 & -43 03 18.4 & 603 & 42132 & 26.31 &            &  \\      
  ctr611  & 13 24 54.78 & -42 54 06.8 & 503 &  4611 & 25.15 &            &  \\      
  ctr803  & 13 24 00.96 & -43 15 16.8 & 530 &  2803 & 25.68 &            &  \\       
  f04p2   & 13 24 30.54 & -43 30 45.9 & 746 &       &       &            &  \\    
  f08p16  & 13 26 03.97 & -43 11 42.4 & 502 &       &       &            &  \\
  f08p46  & 13 25 28.14 & -43 08 15.7 & 482 &  6109 & 24.77 & EMMI\_2307 & 24.87 \\     
  f08p54  & 13 25 12.90 & -43 12 52.0 & 547 &       &       & EMMI\_2361 & 24.64 \\
\hline
\end{tabular}
\label{tab:PN_liter_vels}
\end{table*}

In addition to the list of 1107 PN observed with FLAMES (Table~\ref{tab:PN_ALL_vels}), the 
further 132 PN with radial velocities from the \citet{peng+04PN} catalogue brings the 
total number of PNe with a single velocity component in NGC~5128 with RV measurements to 1239. 
For completeness, in 
Table~\ref{tab:PN_liter_vels} the RV measurements from \citet[][Table 6]{peng+04PN} 
for PN that were not observed with FLAMES are listed. In order to produce a single 
coherent dataset, when combining FLAMES velocities and the literature velocities for PN in 
Table~\ref{tab:PN_liter_vels}, a shift of +7.9 kms$^{-1}$ should be applied to the velocities 
in Table~\ref{tab:PN_liter_vels}. In addition these PNe were cross-identified with the 
\citet{hui+93b} and EMMI imaging catalogues in order to provide 5007\AA\ magnitudes.

Two PN listed in Table~\ref{tab:PN_liter_vels} occur within separations of 1-2$''$
of PN listed in Table~\ref{tab:PN_ALL_vels} and therefore were not paired by the 1$''$ 
matching criterion (see Sec. \ref{sec:astrometry}). These are listed by \citet{peng+04PN} 
as f08p77 and f18p87: f08p77 is within 1.4$''$ of EMMI\_1818, with 
$m(5007\AA)_{\mathrm{EMMI}} = 24.43$, and the velocities differ by 32 kms$^{-1}$; f18p87 
lies within of 1.2$''$ of ID 909 of \citet{hui+93b} (EMMI\_1005 in 
Table~\ref{tab:EMMI_PNcand}), with $m(5007\AA)_{\mathrm{EMMI}} = 24.77$ and
$m(5007\AA)_{\mathrm{Hui}} = 24.98$, and the velocities differ by 23 kms$^{-1}$. 
The histogram of separations of the confirmed PN (Tables~\ref{tab:PN_ALL_vels} and
\ref{tab:PN_liter_vels}) shows a gap between 1.4 and $>$3$''$ suggesting that 
these two PN may be the same objects observed with FLAMES, but the coordinate error 
in the \citet{peng+04PN} catalogue is larger than expected. Given that the 
velocities do not differ by more than twice the combined error of the 
\citet{peng+04PN} and FLAMES GIRAFFE radial velocities, it is suggested that these 
are the same PN. The similarity of their radial velocities is not in itself very 
indicative since both velocities occur within 1$\sigma$ of the peak of the PN velocity 
distribution. In the case of f18p87, the EMMI and Hui 5007\AA\ magnitudes are equal 
within the errors. The surface density of confirmed PNe in the vicinity of both these PNe
is similar at $\sim 0.002$ PN arcsec$^{-2}$, and the probability of finding two objects 
within 1.4$''$ is 1\%. The probability alone is not conclusive to argue for two
separate PN, but, taken with the other evidence, it is suggested that f08p77 and f18p87,
listed in Table~\ref{tab:PN_liter_vels}, are very likely already confirmed PN (listed
in Table~\ref{tab:PN_ALL_vels} as EMMI\_1818 and EMMI\_1005). However, their removal from 
the list of literature PN candidates would entail a revision of the matching criterion 
not motivated by the astrometric errors (c.f. Sect.~\ref{sec:astrometry}), so the two PN 
remain listed in Table ~\ref{tab:PN_liter_vels}.

\section{Results}
\label{sec:results}

\subsection{Efficiency of PN detection}

The use of multi-object spectroscopy to follow-up the on-band / off-band 
detection of planetary nebula candidates presents an opportunity to 
assess the reliability of the detection method. Detection of [O~III] 
emission line sources through imaging depends critically on the presence of 
similar conditions (particularly atmospheric transparency and seeing) in the 
matched on-band and off-band exposures. Considerations also arise from the 
reduction fidelity, such as good astrometry for the placing of matched apertures 
for obtaining the photometry in both bands and flagging of cosmic rays within the 
seeing disk of the point sources. Since only single on-band and
off-band images were used to detect the PN candidates (see Sect. 2.2), cosmic
rays are more likely to be present in the longer [O~III] images which may
not be distinguished from true point sources using the CASU source extraction 
method (contiguous sets of pixels above a threshold). Thus the detected brightness
of weak or non-existent [O~III] emitting sources can be increased by the presence 
of cosmics, biasing the number of emission line candidates in comparison to the  
off-band images which, with shorter exposure time, have a lower density 
of cosmic ray affected pixels. 

In order to quantify the success of the EMMI imaging detection of PNe, the candidates 
selected for spectroscopy were examined. Of the 996 PN candidates selected for 
FLAMES MEDUSA spectroscopy, 592 were found to have a detected line(s), resulting 
in a success rate of 59\% for the selected objects. Comparison of the detected 
and non-detected emission line objects was made in terms of the distribution of 
the assigned EMMI [O~III] magnitudes: there were slightly more fainter 
non-emission line objects, but equally some of the objects with brightest 
m$_{5007\AA}$ did not have emission lines detected. In these latter cases a chance 
superposition of a foreground star, with a spectrum such that the on-off band 
subtraction leaves a positive signal, could perhaps be accountable. 
A good correlation was found 
between EMMI 5007\AA\ magnitude and the detected 5007\AA\ flux converted to magnitude, 
with a correlation coefficient of 0.52 for 398 PN with EMMI magnitudes (probability 
of being uncorrelated is $4 \times 10^{-8}$). Also no correlation was found between 
the location of the fibres within the host galaxy and the non-detection of emission 
lines. Thus there is overall confidence that a 
bright EMMI 5007\AA\ difference image should result in a high [O~III] 5007\AA\ spectral
flux. Examination of the mean spectra for each GIRAFFE exposure for the cases of 
5007\AA\ detected {\it v.} undetected, showed no obvious difference in background 
continuum level or shape. The mean spectra of the non-detected line emitters did 
not show a convincing 5007\AA\ detection for any of the 10 MEDUSA fields 
(Table \ref{tab:GOObs}). 

Seeing and observing conditions do play a strong role in the imaging detection of 
pure emission line sources, as exemplified by the low number of PN detections for 
position FLAMES FIELD6 (see Table \ref{tab:GOObs}), corresponding closely to the EMMI field, 
NGC5128-P7E (Table \ref{tab:EMMI_log}) which was 
affected by thin cirrus and poorer seeing. However the FIELD6 GIRAFFE exposures 
actually had a high detection rate (72\%). It is suggested that the 
effect of cosmic rays on spurious detection of point sources on the longer [O~III] 
images is the major cause for the false emission line sources. 
The approach of using only single detection exposures was
necessitated by the rapid handling of the EMMI data; see Sect.~2.2.
A higher false positive rate occurs for the smaller images with fewer pixels 
in the seeing disk than for larger (poorer seeing) images. This suggestion 
will be confirmed by 
re-reduction of the EMMI imaging with the image pairs carefully combined for each EMMI 
pointing, so that a more robust estimation of the number of PN candidates, 
unaffected by cosmic rays, can be made.

\subsection{Double line PNe}
\label{sec:doublePN}

\begin{table*}
\caption{PNe in NGC~5128 with multiple velocity components. 
}
\begin{tabular}{ l r r r r r r r }
\hline\hline
 Name & $\alpha$~~~~~~ & $\delta$~~~~~~            & Velocity           & Error  & FWHM   & 5007\AA\ flux & [O~III]/H$\beta$ \\
      & (~h ~~m ~~s~) & (~$^\circ$ ~~$'$ ~~~$''$~) & (kms$^{-1}$)   & (kms$^{-1}$) & (kms$^{-1}$) & (Rel.)    &  \\
\hline 

Hui93\_4321 &  13 25 11.69 & -43 08 34.9 & 105.0 &  5.4 &  80.6 &   7.5 &       \\ 	       
            &  13 25 11.69 & -43 08 34.9 & 326.2 &  4.8 &  74.2 &   8.9 &       \\

EMMI\_1105 & 13 25 16.90 & -43 01 05.2 & 486.0 &  0.6 &   47.3 &  60.2 &  12.5 \\ 
           & 13 25 16.90 & -43 01 05.2 & 613.0 &  5.4 &  107.3 &  15.1 &   2.4 \\  
EMMI\_1098  & 13 25 24.59 & -43 00 03.4 & 606.2 &  5.9 &  75.0 &  27.1 &  3.4 \\
            & 13 25 24.59 & -43 00 03.4 & 666.1 &  1.8 &  55.7 &  58.1 &  1.1 \\		
            & 13 25 24.59 & -43 00 03.4 & 749.5 & 16.2 & 108.4 &  15.1 &  3.5 \\

Hui93\_4138  & 13 25 32.66 & -42 58 23.0 & 370.9 &  1.9 &  60.5 &  40.4 &  22.8 \\ 
             & 13 25 32.66 & -42 58 23.0 & 472.8 &  3.2 &  49.4 &  14.6 &   3.7 \\   


Hui93\_4417 & 13 25 10.33 & -42 59 44.3 & 344.6 &  1.3 &  59.2 &  29.8 &	    \\  
            & 13 25 10.33 & -42 59 44.3 & 789.1 &  3.8 &  61.5 &	8.2 &	    \\

Hui93\_525  & 13 25 15.86 & -43 01 56.6 & 344.2 &  2.8 &	43.8 &   8.6 &      \\  
            & 13 25 15.86 & -43 01 56.6 & 789.4 &  1.2 &	55.7 &  41.6 &  16.6 \\  

EMMI\_1918  & 13 26 33.84 & -42 48 47.6 &  53.6 &  6.7 &  71.6 &  6.3 & \\
            & 13 26 33.84 & -42 48 47.6 & 329.1 &  4.4 &  52.4 &  5.2 & \\
                        
%
GC HCH-06 & 13 25 25.48 & -43 01 56.4 & 517.6 &  8.8 &  68.8 &  21.8 & 2.2 \\                          
          & 13 25 25.48 & -43 01 56.4 & 643.3 &  9.1 & 142.6 &  39.0 & 2.3 \\                          

%
GC HGHH-G169 & 13 25 29.44 & -42 58 09.8 & 405.0 &  3.5 &  29.5 &   4.5 & 3    \\
             & 13 25 29.44 & -42 58 09.8 & 689.3 &  0.9 &  46.7 &  56.7 & 6.3	  \\   
\hline
\end{tabular}
\label{tab:PN_doubles}
\end{table*}

\begin{figure}
\centering
\resizebox{\hsize}{!}{
\includegraphics[angle=0]{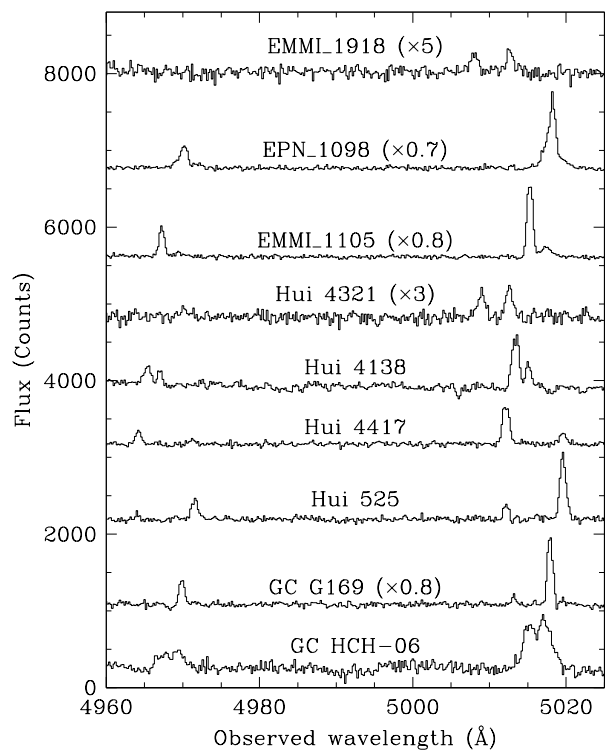}
}
\caption{Spectra of all 9 NGC~5128 PNe, showing double- or triple-lined 
[O~III] profiles, are displayed over the extent of the 4959 and 5007\AA\ region. 
The flux scale is the same for each spectrum (except for the brighter
and fainter fluxes rescaled by factors as indicated), but an 
arbitrary constant background has been applied from each spectrum before plotting.
The two lower spectra are for the double-lined PN in globular clusters.
}
\label{fig:PN_doubles}
\end{figure}

Nine PNe, including two in globular clusters, show evidence 
of a double or complex profile to at least the [O~III] 5007\AA\ line, and often also 
visible on the fainter [O~III]4959\AA\ and H$\beta$ lines. Fig. \ref{fig:PN_doubles}
shows the 4960 - 5025\AA\ region of all these targets and Table 
\ref{tab:PN_doubles} lists the radial velocities, FWHM and relative
fluxes from multiple Gaussian fits to these spectra. 
%
%
A double-lined PN can arise from two main causes: either the asymmetric expansion 
of the ionized shell(s) is resolved in velocity, producing a close double (or perhaps 
triple) profile, or two PN separated along the line of sight defined by the FLAMES 
MEDUSA fibre are detected together. Examination of Fig. \ref{fig:PN_doubles} shows 
obvious examples of both: the PN in GC HCH06 and EMMI\_1098 are fairly unequivocal 
examples of single PN with complex velocity structure, although in both cases the 
line extent is large, challenging the PN assignment; the PN in GC G169, Hui 525, 
4321, 4417 and EMMI\_1918 are fairly unequivocal examples of chance superpositions; 
EMMI\_1105 and Hui 4138 could qualify for either category. We discuss the line 
profiles of these objects in more detail and provide arguments for the above statements.

Given the modest spectral resolution of GIRAFFE with the L3 grating setting
in MEDUSA mode (resolution $\sim$40 kms$^{-1}$,  R=7500 at 5000\AA), the internal expansion 
velocity of a PN must be large for the lines to be resolved. Typically 
the expansion velocity of a PN is between 10 and 70 kms$^{-1}$, as measured by the
FWHM of the [O~III] line, but shows little trend with galaxy type, 
$M_{5007\AA}$, etc, as shown by the kinematic study of extra-galactic PN
by \citet{diaz_men+10}. Extreme objects, such as NGC~6302, a
well-known bipolar Galactic PN, can show line splitting to 
$\sim$80 kms$^{-1}$ and profiles over a range of velocities 
of 180 kms$^{-1}$ \citep{meaburn+walsh80}, 
and the triple profiles of EMMI\_1098 could be an example of such a 
nebula. However it should be noted that the [O~III]/H$\beta$ ratios for the 
three components of EMMI\_1098 are low (Table \ref{tab:PN_doubles}), and not typical 
of high ionization PN, casting some doubt on its identification as a bona-fide
PN. 

The line profile of the PN detected towards the GC HCH-06 PN is a peculiar 
example with line widths of 69 and 143 kms$^{-1}$, a separation of 126 
kms$^{-1}$ (see Table \ref{tab:PN_doubles}), and [O~III]/H$\beta$ 
ratios of 2.2 and 2.3 for these components respectively. 
The high expansion velocity, such broad lines and low [O~III]/H$\beta$ ratio
argues against the identification as a PN and suggests it could perhaps be a nova, 
or a low mass X-ray binary. There is however no detected X-ray source at this 
position in the catalogue of \citet{kraft+01}. This GC is well detected on 
HST WFPC2 images in the programme of C\^{o}t\'{e} et al. (HST Proposal GO6789) in
F555W and F814W filters and is rather elliptical in shape 
\citep[e=0.35 in ][]{holland+99} with a possible loop to the NW visible on the F814W image 
and a partial shell (radius 1.4$''$) to the SE on the F555W image. The detection of 
extended structure in the F555W filter suggests that it could arise from [O~III] 
emission, but clearly detailed observation is required to elucidate the nature of 
this peculiar target.

\citet{minniti+rejkuba02} first detected the PN in G169 and 
this was the first PN detected in the globular cluster of a giant elliptical 
galaxy. \cite{peng+04GCS} in their study of PN and GCs also observed 
this GC but with slightly higher spectral resolution and observed a double peaked 
profile to both [O~III] 4959 \& 5007\AA\ lines. They called into question its
identification as a planetary nebula on the basis of the line splitting of
$\sim$300 kms$^{-1}$ and suggested it might rather be a supernova remnant (SNR).
G169 was targeted in this study to bring higher spectral resolution to bear and
the result is curious. The measured line splitting is 280 kms$^{-1}$ (Tab. 
\ref{tab:PN_doubles}), and both components are narrow, with the redder 
component the stronger one. The redder component is 
within 40 kms$^{-1}$ of the radial velocity of the GC \citep[649 kms$^{-1}$, ][]{woodley+10b},  
and this was confirmed in the FLAMES spectra by measuring the velocity of the
GC H$\beta$ absorption line (see Sec.~\ref{sec:PNinGCs}). However the redder component
appears to be much weaker than the bluer component in the \cite{peng+04GCS} 
observation. The velocity coincidence of the stronger line with the GC velocity is 
within 2.5$\sigma$, based on the combined errors of GC and PN velocities and the 
velocity dispersion for a GC of the absolute V magnitude of G169 (Sec.~\ref{sec:PNinGCs}).

The redder component of G169 has a higher [O~III]/H$\beta$ ratio than the
blue (6.3 v. $\sim$3), although the value for the blue component has large 
errors as the lines are weak. Either the bluer line profile has
strongly declined in brightness (implying very high electron densities for
recombination since only a few years elapsed between the two sets of
independent observations) or this is a displaced emission line target, such that it was
only partially included in the FLAMES MEDUSA 1.2$''$ aperture. The latter
interpretation would favour the +ve component truly being a PN in the
GC population and this is favoured by the [O~III]/H$\beta$ being higher than
for a metal rich HII region. The -ve component might then be a chance line of sight
PN, or HII region, similar to other double-lined PN (c.f. for example 
Hui \#525 in Fig. \ref{fig:PN_doubles}). \citet{peng+04GCS}  also
mention that the [O~III] emission appears to be offset by $1\farcs 4$ to 
the north-east of the GC centre; presumably this refers to
the stronger +ve component but, given the discrepancy, cannot be certain. 
Taken on its own the GIRAFFE spectrum favours two,
presumably unrelated, PN spatially close to each other, but not coincident. An 
extended object, of the same size or larger than the FLAMES fibre
diameter, can be ruled out as then both components, equally strong 
as in the Peng et al.\ observation, should have been observed with FLAMES. 
A typical SNR at 3.8Mpc would be barely resolved even at HST resolution
and is less likely to show narrow line profiles.

EMMI\_1105 rather presents a hybrid case where the separation of the
two line components, at 127 kms$^{-1}$, is very large for a single PN and the
higher velocity component is very broad (FWHM 107 kms$^{-1}$). A high resolution image 
of this target is available on an HST WFC3 UVIS image in F606W filter, taken
in the HST programme GO10597. In the vicinity of the PN position, and within the MEDUSA
fibre, there is a partial filamentary loop structure, assumed to be emission but this
cannot be certain since the passband is broad (but includes H$\alpha$ and [N~II]
emission). It may be that the broad line emission at 630 kms$^{-1}$ arises 
in high velocity extended emission unrelated to the PN. 

EMMI\_1918 is situated in the 
region of the outer NE optical jet with considerable diffuse emission in the vicinity. 
The -ve velocity component is in addition very blue shifted (by 490 kms$^{-1}$) from
the systemic velocity, although both components are fairly narrow and of similar width.
(see Table \ref{tab:PN_doubles} and Fig. \ref{fig:PN_doubles}). Some uncertainty as
to the PN classification of this object is implied and both components may arise
in large scale ionized gas components associated with the emission from the jet
interaction. 

There is a very curious coincidence in radial velocities for the two
PN Hui93\_525 and Hui93\_4417, showing similarity to within the errors for both
components. The relative
strengths of the blue/red [O~III]5007\AA\ strengths is however reversed between the
two sets of profiles. The raw spectra were examined for any instrumental effects
that might have caused this coincidence: the two PN result from different 
fibre positions P2C and P2D (see Table \ref{tab:SVObs}), their $m_{5007A}$ differ
by 0.39 and they are well separated on the sky (distance 146$''$). 

This leaves 4, perhaps 5 objects (if Hui93\_4138 is included) as likely double PN 
resulting from chance line-of-sight superpositions within the MEDUSA GIRAFFE fibre. Given that 
one reason for overluminous PN in the [O~III] PN luminosity function
could be line-of-sight superpositions (\citep{jacoby+90}) , it is interesting to assess whether
4-5 double lined PN is expected in a sample of 1135; the 132 PN with only 
\citet{hui+95} and/or \citet{peng+04PN} velocity measurements (Tab.~\ref{tab:PN_liter_vels}) were not 
included on the grounds that the spectral resolution and signal-to-noise were 
neither sufficient to detect double-lined PN as in Fig. \ref{fig:PN_doubles}. 
A simple model was developed
to generate the surface density and number of detected PN, and to assess how many
double, or triple, detections are expected within a 1.2$''$ FLAMES fibre. Initially the
2MASS image in H band was used to match the surface density of PN, but
given the complex spatial sampling of PN in the FLAMES
data, without systematic area coverage and with varying brightness limits and
restrictions on fibre placement, together with the presence of heavy
extinction of the central dust lane, the measured PN surface density was used
instead. Simple fits along the galaxy major and minor axis were used to generate an
elliptical de Vaucoloueurs distribution with $R_{eff}$=23$'$  and ellipticity 0.45
with surface density set to that of the detected PN (central surface density
extrapolates to 1.4 PN arcsec$^{-2}$). A mask was placed over the core of the galaxy 
to match the size and presence of the dust lane. 

\begin{figure}
\centering
\resizebox{\hsize}{!}{\includegraphics[angle=0]{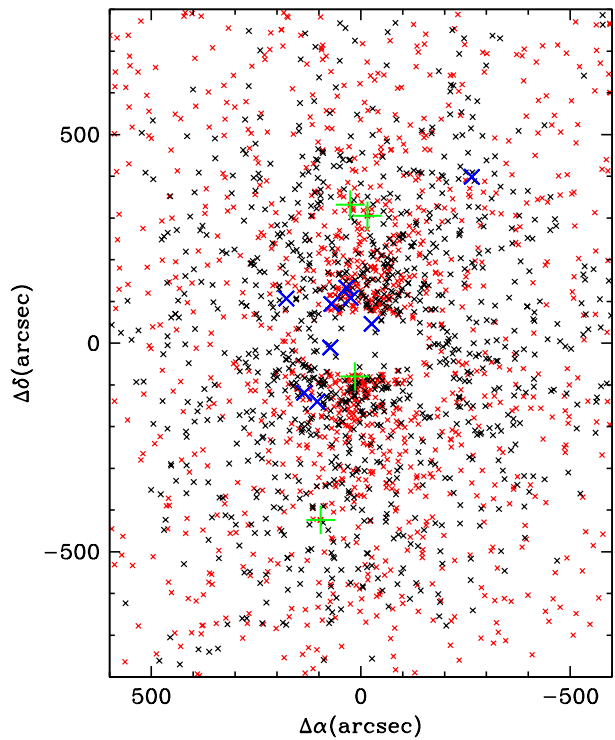}}
\caption{Example of the results of the Monte Carlo simulations for the
number of double-lined PN occurring in the sample of 1135 observed with 
FLAMES. The spatial distribution of PNe is shown with the galaxy rotated 
(by 35$^{\circ}$) with the major axis along the y-axis. The black crosses 
show the observed single-lined PN and the blue
crosses all the double-lined PN (see Fig. \ref{fig:PN_doubles}); the red 
crosses show the simulated single-lined PN. The four double PN resulting from this
particular Monte Carlo simulation are shown by the green pluses. 
No triple-lined PN were detected in this simulation. The central region 
masked out of the simulation corresponds to the extent of the dust lane 
(although a few PNe were detected in this area, including the triple lined
PN EMMI\_1098 and the double lined PN in the globular cluster H6).
}
\label{fig:EMMI_simul_saveO}
\end{figure}

A Monte Carlo code generated the number of observed PN utilising 
the PN surface number density as the probability distribution from
which to sample. Multiple trials were performed (usually 1000) to determine
the mean and standard deviation of the number of double and triple
PN within one FLAMES MEDUSA fibre. Fig. \ref{fig:EMMI_simul_saveO} shows one of the
realizations with the double and triple PN indicated by different coloured symbols.
The average number of double PN detected was 4.8 $\pm$ 2.2 and the average 
number of triple PN produced was always $<$1. 
Variations in the ellipticity and $R_{eff}$ were made but these numbers 
with the standard error bands were typical of those models that did not 
produce too centrally condensed PN surface density realizations.
A rather extreme inclusion of two PN within $1\farcs 2$ was taken, but given 
typical seeing of 1$''$ this may be realistic to detect two sets of
line profiles. The conclusion of these experiments is that the number of 
double lined PN, assumed to be line-of-sight superpositions, is compatible
with the distribution of surface density and the number of PN observed
in this study.

The 5007\AA\ magnitude of the 5 PN accounted by probable line-of-sight
coincidences (excluding the ambiguous case in the globular G169) lie in the 
range 24.2 to 25.6 based on the photometry of \citet{hui+93b} or the calibration 
based on the [O III]5007\AA\ flux against magnitude (Fig. \ref{fig:Corr_FluxMag}). 
If the imaging survey was made
at similar resolution to the MOS spectroscopy (i.e. two objects within $1\farcs 2$ 
are considered as a single object), then none of these double PN would
occur at the bright end of the PNLF in the steeply rising
portion of the distribution \citep[see ][, Fig. 8]{hui+93a}. These
double PN would not therefore significantly affect the 
peak $m_{5007A}$, which plays an important role in the fitting of the PNLF, 
more especially when the number of detected PN is small. 

\subsection{PNe in globular clusters} 
\label{sec:PNinGCs}

\begin{figure}
\centering
\resizebox{\hsize}{!}{
\includegraphics[angle=0]{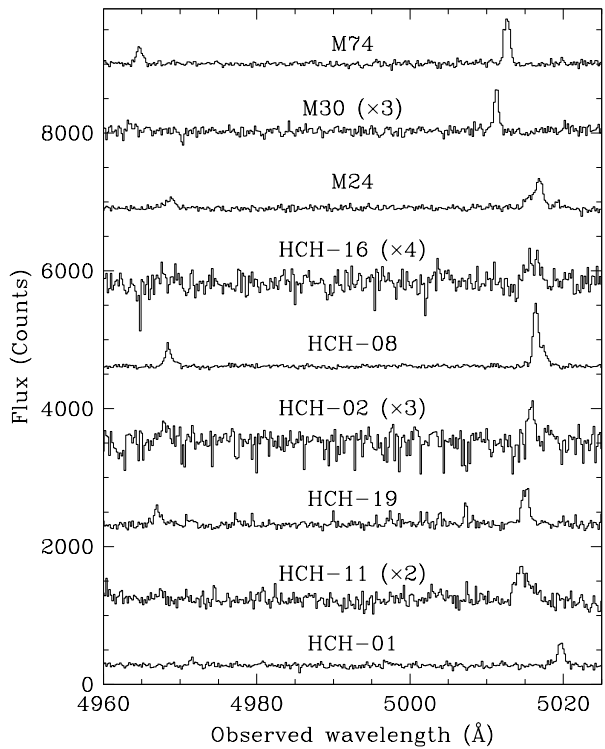}
}
\caption{Spectra of the NGC~5128 globular clusters with
detected [O~III]5007\AA\ line emission, displayed over the 
extent of the 4959 and 5007\AA\ region. The flux scale is the same for 
each spectrum (except for those indicated where the observed signal 
was rescaled for display purposes). For the emission line objects in 
the GCs G169 and HCH-06, see Fig. \ref{fig:PN_doubles}.
}
\label{fig:GCPNs}
\end{figure}

48 globular clusters (GCs) were targeted in the SV observations. Of these, 9, 
excluding HGHH-G169 and HCH-06 (Sect. \ref{sec:doublePN}), 
were found to contain narrow emission lines, suggesting the 
presence of a PN. Table \ref{tab:GCPN} lists the name of the globular cluster, its
coordinates and the PN radial velocity, FWHM, [O~III] relative line strength and
[O~III]/H$\beta$ ratio, where available. Figure \ref{fig:GCPNs} shows the
[O~III]4959,5007\AA\ emission line region of these GCs. The [O~III]/H$\beta$ 
ratio is a useful discriminant between a bona fide PN, with higher values of this
ratio, and an HII region or nebula ionized by a cooler star or perhaps shock ionized,
with generally lower values. The HII regions observed in the dust lane of NGC~5128 
for example have [O~III]/H$\beta$ ratios around 0.60 \citep{moellenhoff81, phillips81};
this value can be contrasted with the distribution of [O~III]/H$\beta$ ratios
in Fig. ~\ref{fig:Hist_o3hb}.

\begin{table*}
\caption{PNe in NGC~5128 globular clusters (excluding two with multiple velocity components, 
see Tab.~\ref{tab:PN_doubles}).
}
\begin{tabular}{ l r r r r r r r }
\hline\hline
 GC & $\alpha$~~~~~~  & $\delta$~~~~~~ & Velocity & Error  & FWHM   & 5007\AA\ flux & [O~III]/H$\beta$ \\   
    & (~h ~~m ~~s~) & (~$^\circ$ ~~$'$ ~~~$''$~) & (kms$^{-1}$)   & (kms$^{-1}$) & (kms$^{-1}$) & (Rel.)   &                  \\   
\hline 

%

%
HCH-01  & 13 25 16.22 & -42 59 43.4 & 795.5 &   2.4 &  57.6 &   15.4 &  1.0 \\

%
HCH-02   & 13 25 16.68 & -43 02 08.7 & 565.9 &   4.9 &  62.9 &   8.8 &   \\                          

%
HCH-08   & 13 25 26.95 & -43 00 01.3 & 604.0 &   0.9 &  64.4 &  73.3 & 0.9 \\

%
HCH-11  & 13 25 27.49 & -43 01 52.0 & 492.8 &  10.7 & 125.6 &   22.3 &   \\                        

%
HCH-16   & 13 25 30.30 & -42 59 34.8 & 577.9 &   6.6 & 140.7 &  28.6 &  \\  

%
HCH-19  & 13 25 31.75 & -43 00 33.7 & 548.8 &   3.8 &    85.2 &  23.3 &  1.8 \\  

%
M24  & 13 25 19.84 & -43 00 54.0 & 616.8 &   2.7 & 103.6 &  29.2 &  3.1 \\                          

%
M30  & 13 25 40.45 & -43 02 51.3 & 290.3 &   1.5 &  41.9 &  15.6 &  1.0 \\ 

%
M74  & 13 25 18.52 & -43 01 16.0 & 639.7 &   9.7 & 122.3 &  20.2 & 11.3 \\ 
 \hline
%
%
\end{tabular}
\label{tab:GCPN}
\end{table*}

\begin{table*}
\caption{Comparison of GC and GC PNe radial velocities}
\label{tab:GCPNvels}
\centering 
\begin{tabular}{ l l r r r r l r r l }
\hline\hline

 GC &  Alt. & $V$(GC) & Error  & $V$(PN) & Error & $M_{T1}$ & $|\Delta$($V_{PN} - V_{GC})|$ &
 $3 \times \sigma_{T}$ & PN \& GC \\
    &              & (kms$^{-1}$) & (kms$^{-1}$) & (kms$^{-1}$) & (kms$^{-1}$) & (mag.) & (kms$^{-1}$) & (kms$^{-1}$) & related \\   
\hline 

HCH-01$^{1}$    & GC0168$^{2}$ &  623 &  42 & 796 &  2 & 18.6$^{W1}$ & 173 & 138 & No \\
HCH-02$^{1}$    & GC0171$^{2}$ &  291 &  20 & 566 &  5 & 17.6$^{*}$ & 275 &  75 & No \\
HCH-06$^{1,Y}$  &              &  680 &  51 & 643 &  9 & 18.9$^{*}$ &  37 & 101 & Yes \\
HCH-06$^{1,Z}$  &              &  680 &  51 & 518 &  9 & 18.9$^{*}$ & 162 & 101 & No \\
HCH-08$^{1}$    &              &  637$^{X}$ &     & 604 &  1 & 19.6$^{*}$ &  33 & 60 & Yes? \\
HCH-11$^{1}$    &              &  641 &  17 & 493 & 11 & 18.0$^{*}$ & 148 &  75 & No \\
HCH-16$^{1}$    &              &  443 &  18 & 578 &  7 & 18.0$^{*}$ & 135 &  73 & No \\
HCH-19$^{1}$    &              &      &     & 549 &  4 & 19.5$^{*}$ &     &     &  ? \\  
HGHH-G169$^{3,Y}$ &              &  649$^{W2}$ &  16 & 689 &  1 & 19.0$^{*}$ & 40 & 56 & Yes \\
M24$^{4}$       &              &  467 & 175 & 619 &  3 & 21.4$^{*}$ & 232 & 525 & Yes? \\
M30$^{4}$       &              &  792 &  43 & 290 &  2 & 20.2$^{*}$ & 502 & 166 & No \\
M74$^{4}$       & GC0182$^{2}$ &  651 &  64 & 640 & 10 & 19.4$^{*}$ &  11 & 195 & Yes \\ 
\hline
\end{tabular}
\tablefoot{
 $^{1}$ HCH = \citet{holland+99} \\
 $^{2}$ \citet{woodley+07} \\
 $^{3}$ HGHH = \citet{harris+92} \\
 $^{4}$M = \citet{minniti+04} with coordinates from \citet{kraft+01} \\ 
 $^{Y}$ +ve component of double emission line only \\
 $^{Z}$ -ve component of double line only \\
 $^{*}$ Adopted from observed V or observed R mag \\
 $^{X}$ Cross correlation failed - estimated by eye \\
 $^{W1}$ T1 photometry from \citet{woodley+07} \\
 $^{W2}$ Radial velocity from \citet{woodley+10a} \\
}
\end{table*}

In order to determine if the PN and the GC are related, the most direct method 
is to compare their radial velocities. The velocities of some of these GCs are 
available in the literature \citep{peng+04GCcat, woodley+07, woodley+10b}. 
Velocities can also be determined by cross-correlation using the GIRAFFE spectra. 
Given that no standard stars were taken for this purpose with the FLAMES in 
MEDUSA mode, a cross-correlation template was formed by averaging the three 
highest signal-to-noise GC spectra (HGHH-C6, HCH-21 and HGHH-C4),
%
%
and shifting the other two spectra to that of HGHH-C4. The radial velocity of this 
GC was measured by \citet{woodley+07} as 689 $\pm$ 16 kms$^{-1}$ and 
so was taken as the template velocity. Cross correlating the other GC spectra against
this template, and comparing the measured velocities with those from \citet{woodley+07, 
woodley+10b} for 17 GCs enabled a modified template 
velocity to be determined which better matched, on average, the velocities from the
literature, and resulted also in consistent velocity for HGHH-C4 within the error. 
All cross-correlation measurements were preformed using the IRAF {\it fxcor} task.  

For a PN to be directly associated with the GC, i.e., bound by the
GC gravitational potential and not a chance line-of-sight coincidence,  
its radial velocity must be close to that of the GC. \citet{jacoby+13} 
surveyed many GCs in M31 for evidence of associated PN through the 
detection of the [O~III]5007\AA\ line and suggested a criterion 
$\Delta v ^{<}_{\sim} 3 \sigma_{T}$, where $\Delta v$ is the absolute 
difference of the PN and GC velocities and $\sigma_{T}$ is the quadratic 
sum of the errors on the PN and GC velocities and the internal velocity 
dispersion of the GC. Table \ref{tab:GCPNvels} compares the $\Delta v$ and 
$\sigma_{T}$ values, except for HCH-08 and HCH-19. These two
clusters had noisy continuum spectra and no other velocity was found for them
in the literature.  The velocity dispersion of each 
GC, $\sigma_{GC}$, was estimated from the absolute $T_{1}$ 
magnitude following \citet{jacoby+13} Eqn. 1. If $T_{1}$
was not available, the $R$ magnitude was employed taking the mean $(R-T_{1})$ 
= -0.080 from the \citet{woodley+07} catalogue; where the 
$R$ mag was not available, it was estimated based on the $V$ mag and a constant 
$(V-R)$ colour of 0.547 from the mean of Woodley's GC photometry. A single 
value of reddening of $A_V$ of 0.40 from \citet{schlegel+98} was 
adopted. The values of $\sigma_{GC}$ ranged from 3 to 15 kms$^{-1}$. The 
largest contributor to $\sigma_{T}$ in Table \ref{tab:GCPN} mostly arises 
from the uncertainty on the GC radial velocity. 

Two GCs show convincing evidence for a bound PN (i.e. $\Delta v < 3 \sigma_{T}$): 
M74 and HGHH-G169. The velocity of HCH-06 associates it with the positive 
velocity component of the emission source, but the nature of this source suggests 
it may not be a PN. Another GC, M24, shows possible evidence for association with a PN, 
although the large error on the GC velocity renders this much less convincing. 
Of these three candidate 
GCs with associated PN, H$\beta$ could be measured once the template spectrum, shifted 
to the radial velocity of the GC, had been subtracted. G169 was discussed extensively 
in Sect. 6.2 and the [O~III]/H$\beta$ ratio of the component closer to the GC velocity 
shows a fairly high ratio, arguing for a PN; the lower velocity component shows a lower
ratio so may not be a PN.  M74 has a detected H$\beta$ and the [O~III]/H$\beta$
ratio is 11 (Table \ref{tab:GCPN}), strongly suggesting an association with a PN. 
M24 shows a low [O~III]/H$\beta$ ratio (Table \ref{tab:GCPN}); although not conclusively 
ruling out a PN (younger PN with a lower temperature central star or a high density 
shell, can display lower values of [O~III] flux), this casts some doubt on the 
emission source being a PN. M24 is additionally interesting 
in that it shows a [O~III] 5007\AA\ line profile with a blue extension, which can be 
fitted by two components with a velocity separation of 85 kms$^{-1}$. This GC exhibits
an X-ray detection with Chandra of $6.7\times10^{36}$ ergs s$^{-1}$ 
in the 0.4-10 keV band in \citet{kraft+01}. 

\subsection{Emission line spectra in the north-eastern filaments}
\label{sec:skyfibres}

In the two FLAMES observing runs, typically $\sim 15-20$ fibres were allocated
in each setup to assumed empty sky regions. In a few cases emission line spectra 
were detected in the sky fibre.  Since the Period 73 
open time run for Fields 7 and 8 configurations (Table \ref{tab:GOObs}) were located in the region of the 
north-eastern filaments present along the radio and X-ray jet \citep{graham98}, by chance a few sky fibres 
contained emission lines. Close inspection showed that these are indeed connected to the ionized gas in the filaments. 

In Field 7 three sky fibres fell on top of the filaments:
\begin{enumerate}
\item  The sky fibre \#66 had RA=13 26 26.44, Dec=-42 50 10.9, and is located 
in the outer filament of the NE halo, roughly between slits A and C of \citet{graham98}. 
The measured velocities in \citet{graham98} are 240-355 kms$^{-1}$ 
in that position (slits A and C, respectively) while 410-416 kms$^{-1}$ was measured 
with FLAMES. 
\item The sky fibre \#182 had RA=13 26 20.87 and Dec=-42 54 41.5. This is to the north-east of the inner filament.
The emission was only very weakly detected in the 5007\AA\ line in one of the two exposures (Field7\_1) corresponding
to a velocity of 591 kms$^{-1}$. However, this detection is very uncertain, and could be affected by a 
cosmic ray, given the lack of detection in exposure Field7\_2.
\item The sky fibre \#186 had RA=13 26 05.01 and Dec=-42 56 57.2. This by chance fell on top of knot C in 
the inner filament observed by \citet{graham+price81}. Their [O~III] (5007+4959)/H$\beta$
ratio is 14, while a value of 8.0 was measured on the GIRAFFE spectrum (5007\AA/H$\beta$ = 5.9). 
The velocity of 252 kms$^{-1}$ measured with FLAMES is somewhat different from the 
355 kms$^{-1}$ measured by \citet{graham98} for knot C.
\end{enumerate}

In FIELD8 only one sky fibre contained emission lines, and this was exactly the 
same position as the sky fibre \#186 from the configuration of FIELD7. 


\section{Discussion}
\label{sec:Discussion}

The FLAMES multi-object spectroscopy of 1135 emission line sources 
presented here provides a rich database for studies of the planetary 
nebula population in NGC~5128. As the nearest large early-type galaxy, 
both the number of PNe and their detailed distribution, both 
spatially and in velocity, are an excellent phase-space 
probe of the gravitational potential of this galaxy. 
Previous studies of the luminous and dark matter potentials as
revealed by 431 PNe (\citet{hui+95}) and 780 PNe (\citet{peng+04PN}) 
with the addition of 340 globular cluster velocities 
(\citet{woodley+07}) can now be further advanced with the 
availability of the 1107 accurate velocities for PNe with
single velocity components (Table 
\ref{tab:PN_ALL_fluxes}) and the addition of another 132 from 
previous studies (placed on a consistent velocity scale, 
Table \ref{tab:PN_liter_vels}). 

\begin{figure}
\centering
\resizebox{\hsize}{!}{\includegraphics[angle=-90]{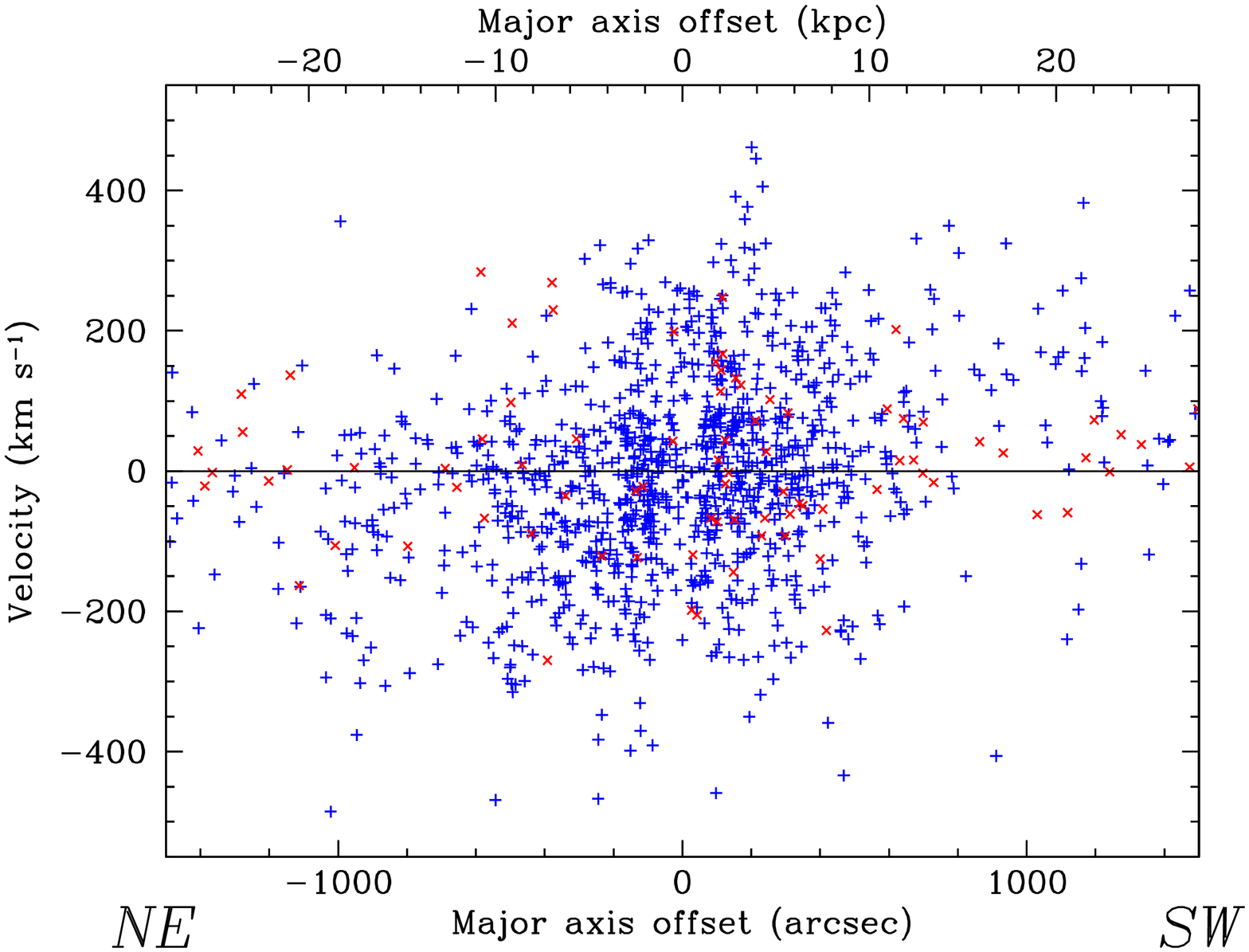}}
\resizebox{\hsize}{!}{\includegraphics[angle=-90]{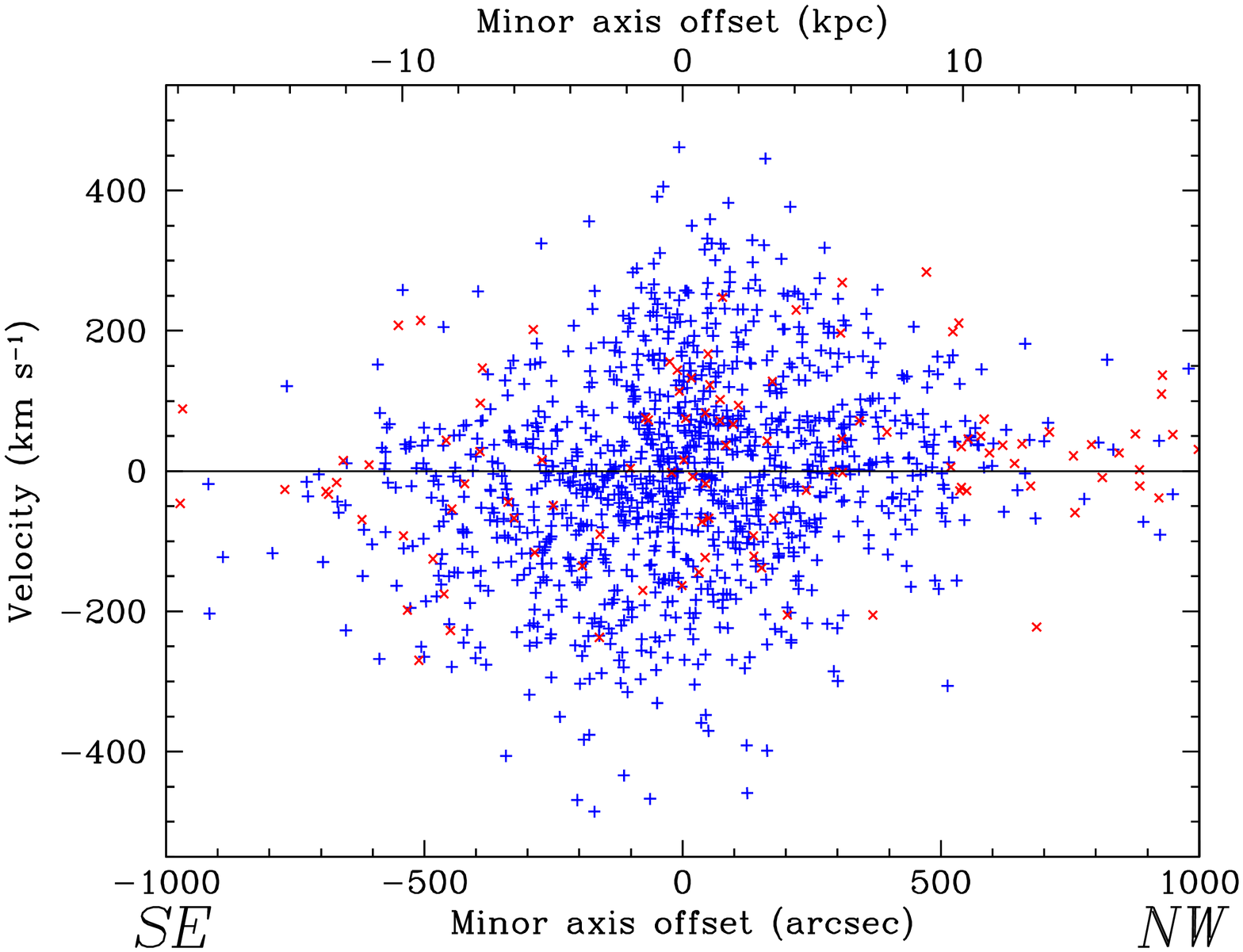}}
\caption{
Upper panel: Rest frame velocity of all the PN in NGC~5128 with an observed 
velocity, plotted versus position along the major axis (PA 35$^{\circ}$). 
The systemic velocity of the galaxy is taken as 539 kms$^{-1}$. Blue points are from the FLAMES SV 
and open time observations; red points for the literature values from 
Table \ref{tab:PN_liter_vels}. \newline
Lower: As for the upper panel, but plotted against offset along the minor axis (PA 125$^{\circ}$).
}
\label{fig:MajMinVels}
\end{figure}

Figure \ref{fig:MajMinVels} displays all the PN velocities as a 
function of their offsets along the major and minor axis (PA of major 
axis taken as 35$^\circ$ from \citet{hui+95}). From 
the mean velocity of the 1135 FLAMES PN velocities, a systemic 
velocity of 539 kms$^{-1}$ for NGC~5128 is adopted and used 
to plot the local offset velocity in Fig. \ref{fig:MajMinVels}. 
This value of the systemic velocity can be compared with values of 
541 kms$^{-1}$ for the 431 PNe in Hui et al. (1995) and 538 
kms$^{-1}$ from absorption line spectra (\citet{wilkinson+86}. 
Red points on Fig. \ref{fig:MajMinVels} correspond to  
literature PN velocities from \citet{hui+95} and \citet{peng+04PN}. 
The two dimensional velocity field can be modelled from these data 
using techniques such as adaptive Gaussian kernel smoothing 
(\citet{coccato+09}) or fitting of non-parametric kinematic 
profiles using maximum-likelihood (\citet{cortesi+11} et al). 
\citet{peng+04PN} found evidence for a triaxial potential. 
Jeans equation modelling has been applied to the rotation and 
velocity dispersion profile derived from PN datasets, such 
as by \citet{hui+95} and \citet{peng+04PN} to estimate the 
mass and mass-to-light ratio of the galaxy and the halo in 
particular to study the presence of dark matter. The 
improvements in the size of the PN velocity database and, 
more particularly, the accuracy of the radial velocities 
measured with FLAMES, will allow more extensive modelling 
of NGC~5128.

Another application which can be refined is 
the search for sub-groups of PNe revealed by their 
phase-space proximity as conducted by \citet{woodley+11}
on the basis of the \citet{peng+04PN} velocity data.  
\citet{coccato+13} compared the halo kinematics of several 
early-type galaxies, including NGC~5128, on the basis of 
the literature GC and PNe velocities. Whilst the PNe show 
clear evidence of disk-like rotation, this is not found for 
the GCs. \citet{coccato+13} found no evidence for the presence 
of sub-structures, unlike the case of NGC 1399, the central galaxy in 
the Fornax cluster. More radial velocities of the candidate 
GCs in NGC~5128, such as the 800 candidates found from B- and R-band 
imaging by \citet{harris+12}, will provide a number of velocity 
probes (PNe and GCs) exceeding 2000 in NGC 5128. Of course 
spectroscopic follow-up of the 1118 PN candidates, 1060 found by
the EMMI imaging (Sect. \ref{sec:imaging}) and the rest from 
literature surveys, and deeper 
imaging surveys have the potential to confirm many more PNe, 
particularly in the outer regions where the density of probes 
is lower and their effect on the derivation of the rotation 
curve is more critical.

A survey designed to find PNe as point source line emitters 
in a galaxy is subject to a number of contaminants. In the 
case of NGC~5128, the many H\,II regions in the area 
of the central dust can provide a possible contaminant. 
However at the distance of 3.8 kpc, most of these should be 
resolved, except for compact HII regions, and can thus 
be effectively discriminated from PNe, that strictly appear 
as point sources at the distance of NGC~5128. However some 
small brighter knots in extended HII complexes may be 
distinguished in point source emission line surveys. A case 
in point was the emission lines detected in the FLAMES 
fibres in the region of the emission filaments, associated 
with the radio jet, featured in Sect. \ref{sec:skyfibres}. 
However these were detected by accident in the offset sky 
fibres. The [O~III]/H$\beta$ ratio can be a useful discriminant 
of PNe ionized by very hot stars, in contrast to H\,II regions
with cooler stars. However in NGC~5128 the emission filaments
associated with the jet can show an [O~III]/H$\beta$ ratio
in the same range as for the PNe (for example sky fibre \#186 
with a ratio of 5.9 - Sect.\ref{sec:skyfibres}), and
also supernova remnants and H\,II regions, so this line 
ratio is not a very reliable discriminant of PNe in this 
particular case. 

Some of the PN detections, particularly those at very low and
very high velocities, may be emission knots associated with the 
radio jet, such as the spectra of the offset sky positions presented in 
Sect.~\ref{sec:skyfibres}. There are four emission line detections with 
radial velocity $<$ 100kms$^{-1}$ ($>$440 kms$^{-1}$ with respect to systemic) -- 
EMMI\_1918, EMMI\_2033, EMMI\_901 and EMMI\_1439 -- and these 
may not be PNe. EMMI\_1918 and EMMI\_2033 both lie in the region of the
outer and inner NE radio jet respectively; EMMI\_901 lies to
the north of the nucleus but not in the direction of the radio jet, 
whilst EMMI\_1439 lies WSW of the nucleus. EMMI\_1439 has a broader 
line (91 kms$^{-1}$)) than the mean of all detected PNe 
(FWHM 54.2 kms$^{-1}$). Only EMMI\_2033 has a measured 
low [O~III]/H$\beta$ ratio (3.5) casting some doubt on its 
validity as a PN; EMMI\_1918  and EMMI\_1439 have no 
H$\beta$ line detected. Since EMMI\_901 and EMMI\_1439 do
not appear to be related to the radio jet they could perhaps be genuine
high velocity PNe, or even high velocity clouds. 

Optical line emission from the redshifted lobe 
of the radio jet, which presumably extends to the SW of the 
nucleus, has never been detected. There are a few detected 
line emitters at high positive velocity at an offset of 
about 4kpc along the major axis (see Fig. 
\ref{fig:MajMinVels}) which could be plausibly considered 
as arising from jet emission. Six PNe have offset velocity 
$> +360$ kms$^{-1}$ (all from \citet{hui+95} -- 2207, 530, 
519, 4249, 42124, 42106) and occur to the SW of the nucleus,
but none show any sign of being other than {\it bona fide} PNe based 
on their line widths and [O~III]/H$\beta$ ratios (only 42124 
did not have H$\beta$ detected). Of these PNe only 2207 
occurs close to the major axis (minor axis offset 89$''$).  

Another contaminant which can occur is high redshift 
emission line galaxies, such as the [O~II] line at $z \sim 
0.34$, the C~III] line at $z \sim 1.63$ or even 
Ly-$\alpha$ at $z \sim 3.1$ (for the latter compare 
\citet{kudritzki+99}). For [O~II], the 3727\AA\ line is a 
doublet with separation 2.8\AA, so that double line 
detections with a  separation of 225 kms$^{-1}$ could 
be detected from $z \sim 0.34$ starburst galaxies. None of 
the double or complex line profile targets in Table. 
\ref{tab:PN_doubles} have this separation and all, except 
EMMI\_1918, have the [O~III]4959\AA\ component detected 
so cannot be high redshift interlopers.

The number of globular clusters showing emission lines 
(Tab.~\ref{tab:GCPN}), suggested as arising from PNe, is high --- 
11 out of the 48 observed with FLAMES. The analysis presented 
in Table \ref{tab:GCPNvels} suggests that only three of 
these are convincingly associated with the GC, with two 
others possibly associated. Comparing to the number of 
PNe known in Galactic GCs (4 PNe out of 130 surveyed) and 
M31 (3/274), see \citet{jacoby+13}, this number is quite 
high. Extrapolating to the estimated total number of about 
2000 GCs in NGC~5128, suggests the surprizing number of 
around 200 PNe in GCs. A more extensive and 
representative survey of line emission in the GCs of 
NGC~5128 is clearly required to investigate if
the proportion of GCs with PNe is unusual compared to
other galaxies, and if there is any dependence on the
GC properties (e.g., colour, luminosity, metallicity, etc).

\section{Conclusions}
\label{sec:Conclusions}
A double pronged study of the planetary nebulae in 
NGC~5128 is presented, consisting of an on-band/off-band 
imaging survey with EMMI to increase the number of PN 
candidates and a FLAMES Medusa spectroscopic follow-up 
of a fraction of these candidates, as well as PNe in NGC~5128 
from the literature. The FLAMES GIRAFFE spectrograph is very well 
suited to the spectroscopic confirmation of the PNe in 
NGC~5128, and the emission line velocity was measured for 
1135 objects with typical errors of a few kms$^{-1}$. 
486 of these PNe are newly confirmed based on the 
measured emission lines.
For most of the PNe both the 4959 and 5007\AA\ components of 
the [O~III] doublet were detected and for more than half (57\%) 
of the PN spectra, the H$\beta$ line 
was also detected. These detections of other lines allow 
confirmation that the targets are PNe, that they reside 
in NGC~5128 and are not higher redshift contaminants. 

Such a large collection of spectra of emission line 
objects is sure to uncover some anomalies and the 
targets with double or complex line profiles and the
11 emission line objects spatially coincident (within 
the Medusa aperture of 1.2$''$) with globular clusters 
are highlighted. About half of the double lined objects 
could arise from line of sight coincidences of two PNe 
within the Medusa fibre and it is shown that this number 
is reasonably consistent with the number expected for 
the observed sample of PNe in this galaxy. The number of 
emission line objects associated with GCs is not 
insubstantial and suggests that many such 
targets could be present in the whole galaxy based on 
its GC population.

Catalogues in machine readable form are presented of 
all the PN candidates and, for those spectrally observed, 
radial velocity, line width and, where available, 
[O~III]/H$\beta$ ratio, are tabulated. In addition 
radial velocities for a further 132 PNe from the literature, 
not covered in this work, are listed and the offset to
the FLAMES velocities is provided. We 
encourage exploitation of these high quality data by the 
community for modelling the gravitational potential 
of NGC~5128 and the investigation of the past history of 
assembly of the galaxy.



\begin{acknowledgements}

Laurie Shaw participated in the NTT observations and we thank him 
for his excellent support. We thank the Paranal staff and the FLAMES SV 
team, led by Francesca Primas,
for their help during the science verification. The Paranal Science
Operations and the User Support Department are warmly thanked for
their contributions to the successful service mode run 073.B-0434(B). The La Silla
staff are thanked for their assistance with the visitor mode NTT programme 073.B-0434(A).
We are also grateful to the CASU team for their rapid processing of the EMMI imaging
datasets. George Jacoby kindly read the manuscript and provided valuable comments. 
\end{acknowledgements}

\bibliographystyle{aa}
\bibliography{mybiblio}

\begin{thebibliography}{62}
\expandafter\ifx\csname natexlab\endcsname\relax\def\natexlab#1{#1}\fi

\bibitem[{{Beasley} {et~al.}(2008){Beasley}, {Bridges}, {Peng}, {Harris},
  {Harris}, {Forbes}, \& {Mackie}}]{beasley+08}
{Beasley}, M.~A., {Bridges}, T., {Peng}, E., {et~al.} 2008, \mnras, 386, 1443

\bibitem[{{Blecha} {et~al.}(2000){Blecha}, {Cayatte}, {North}, {Royer}, \&
  {Simond}}]{Blecha+00}
{Blecha}, A., {Cayatte}, V., {North}, P., {Royer}, F., \& {Simond}, G. 2000, in
  Society of Photo-Optical Instrumentation Engineers (SPIE) Conference Series,
  Vol. 4008, Optical and IR Telescope Instrumentation and Detectors, ed.
  M.~{Iye} \& A.~F. {Moorwood}, 467--474

\bibitem[{{Coccato} {et~al.}(2013){Coccato}, {Arnaboldi}, \&
  {Gerhard}}]{coccato+13}
{Coccato}, L., {Arnaboldi}, M., \& {Gerhard}, O. 2013, \mnras, 436, 1322

\bibitem[{{Coccato} {et~al.}(2009){Coccato}, {Gerhard}, {Arnaboldi}, {Das},
  {Douglas}, {Kuijken}, {Merrifield}, {Napolitano}, {Noordermeer},
  {Romanowsky}, {Capaccioli}, {Cortesi}, {de Lorenzi}, \&
  {Freeman}}]{coccato+09}
{Coccato}, L., {Gerhard}, O., {Arnaboldi}, M., {et~al.} 2009, \mnras, 394, 1249

\bibitem[{{Combes} {et~al.}(2007){Combes}, {Young}, \& {Bureau}}]{combes+07}
{Combes}, F., {Young}, L.~M., \& {Bureau}, M. 2007, \mnras, 377, 1795

\bibitem[{{Cortesi} {et~al.}(2011){Cortesi}, {Merrifield}, {Arnaboldi},
  {Gerhard}, {Martinez-Valpuesta}, {Saha}, {Coccato}, {Bamford}, {Napolitano},
  {Das}, {Douglas}, {Romanowsky}, {Kuijken}, {Capaccioli}, \&
  {Freeman}}]{cortesi+11}
{Cortesi}, A., {Merrifield}, M.~R., {Arnaboldi}, M., {et~al.} 2011, \mnras,
  414, 642

\bibitem[{{Davis} {et~al.}(2011){Davis}, {Alatalo}, {Sarzi}, {Bureau}, {Young},
  {Blitz}, {Serra}, {Crocker}, {Krajnovi{\'c}}, {McDermid}, {Bois}, {Bournaud},
  {Cappellari}, {Davies}, {Duc}, {de Zeeuw}, {Emsellem}, {Khochfar},
  {Kuntschner}, {Lablanche}, {Morganti}, {Naab}, {Oosterloo}, {Scott}, \&
  {Weijmans}}]{davis+11}
{Davis}, T.~A., {Alatalo}, K., {Sarzi}, M., {et~al.} 2011, \mnras, 417, 882

\bibitem[{{Dekker} {et~al.}(1986){Dekker}, {Delabre}, \& {Dodorico}}]{emmi}
{Dekker}, H., {Delabre}, B., \& {Dodorico}, S. 1986, in Instrumentation in
  astronomy VI; Proceedings of the Meeting, Tucson, AZ, Mar. 4-8, 1986. Part 1
  (A87-36376 15-35). Bellingham, WA, Society of Photo-Optical Instrumentation
  Engineers, 1986, p. 339-348., ed. D.~L. {Crawford}, 339--348

\bibitem[{{Dufour} {et~al.}(1979){Dufour}, {Harvel}, {Martins}, {Schiffer},
  {Talent}, {Wells}, {van den Bergh}, \& {Talbot}}]{dufour+79}
{Dufour}, R.~J., {Harvel}, C.~A., {Martins}, D.~M., {et~al.} 1979, \aj, 84, 284

\bibitem[{{Ferrarese} {et~al.}(2007){Ferrarese}, {Mould}, {Stetson}, {Tonry},
  {Blakeslee}, \& {Ajhar}}]{ferrarese+07}
{Ferrarese}, L., {Mould}, J.~R., {Stetson}, P.~B., {et~al.} 2007, \apj, 654,
  186

\bibitem[{{Graham}(1979)}]{graham79}
{Graham}, J.~A. 1979, \apj, 232, 60

\bibitem[{{Graham}(1998)}]{graham98}
{Graham}, J.~A. 1998, \apj, 502, 245

\bibitem[{{Graham} \& {Price}(1981)}]{graham+price81}
{Graham}, J.~A. \& {Price}, R.~M. 1981, \apj, 247, 813

\bibitem[{{Harris}(2010)}]{harris10}
{Harris}, G.~L.~H. 2010, \pasa, 27, 475

\bibitem[{{Harris} {et~al.}(1992){Harris}, {Geisler}, {Harris}, \&
  {Hesser}}]{harris+92}
{Harris}, G.~L.~H., {Geisler}, D., {Harris}, H.~C., \& {Hesser}, J.~E. 1992,
  \aj, 104, 613

\bibitem[{{Harris} {et~al.}(2012){Harris}, {G{\'o}mez}, {Harris}, {Johnston},
  {Kazemzadeh}, {Kerzendorf}, {Geisler}, \& {Woodley}}]{harris+12}
{Harris}, G.~L.~H., {G{\'o}mez}, M., {Harris}, W.~E., {et~al.} 2012, \aj, 143,
  84

\bibitem[{{Harris} {et~al.}(2010){Harris}, {Rejkuba}, \& {Harris}}]{harris+10}
{Harris}, G.~L.~H., {Rejkuba}, M., \& {Harris}, W.~E. 2010, \pasa, 27, 457

\bibitem[{{Harris} {et~al.}(1988){Harris}, {Harris}, \& {Hesser}}]{harris+88}
{Harris}, H.~C., {Harris}, G.~L.~H., \& {Hesser}, J.~E. 1988, in IAU Symposium,
  Vol. 126, The Harlow-Shapley Symposium on Globular Cluster Systems in
  Galaxies, ed. J.~E. {Grindlay} \& A.~G.~D. {Philip}, 205--214

\bibitem[{{Hesser} {et~al.}(1986){Hesser}, {Harris}, \& {Harris}}]{hesser+86}
{Hesser}, J.~E., {Harris}, H.~C., \& {Harris}, G.~L.~H. 1986, \apjl, 303, L51

\bibitem[{{Hesser} {et~al.}(1984){Hesser}, {Harris}, {van den Bergh}, \&
  {Harris}}]{hesser+84}
{Hesser}, J.~E., {Harris}, H.~C., {van den Bergh}, S., \& {Harris}, G.~L.~H.
  1984, \apj, 276, 491

\bibitem[{{Holland} {et~al.}(1999){Holland}, {C{\^o}t{\'e}}, \&
  {Hesser}}]{holland+99}
{Holland}, S., {C{\^o}t{\'e}}, P., \& {Hesser}, J.~E. 1999, \aap, 348, 418

\bibitem[{{Hui} {et~al.}(1993{\natexlab{a}}){Hui}, {Ford}, {Ciardullo}, \&
  {Jacoby}}]{hui+93a}
{Hui}, X., {Ford}, H.~C., {Ciardullo}, R., \& {Jacoby}, G.~H.
  1993{\natexlab{a}}, \apj, 414, 463

\bibitem[{{Hui} {et~al.}(1993{\natexlab{b}}){Hui}, {Ford}, {Ciardullo}, \&
  {Jacoby}}]{hui+93b}
{Hui}, X., {Ford}, H.~C., {Ciardullo}, R., \& {Jacoby}, G.~H.
  1993{\natexlab{b}}, \apjs, 88, 423

\bibitem[{{Hui} {et~al.}(1995){Hui}, {Ford}, {Freeman}, \& {Dopita}}]{hui+95}
{Hui}, X., {Ford}, H.~C., {Freeman}, K.~C., \& {Dopita}, M.~A. 1995, \apj, 449,
  592

\bibitem[{{Israel}(1998)}]{israel98}
{Israel}, F.~P. 1998, \aapr, 8, 237

\bibitem[{{Jacoby} {et~al.}(2013){Jacoby}, {Ciardullo}, {De Marco}, {Lee},
  {Herrmann}, {Hwang}, {Kaplan}, \& {Davies}}]{jacoby+13}
{Jacoby}, G.~H., {Ciardullo}, R., {De Marco}, O., {et~al.} 2013, \apj, 769, 10

\bibitem[{{Jacoby} {et~al.}(1990){Jacoby}, {Ciardullo}, \& {Ford}}]{jacoby+90}
{Jacoby}, G.~H., {Ciardullo}, R., \& {Ford}, H.~C. 1990, \apj, 356, 332

\bibitem[{{Kainulainen} {et~al.}(2009){Kainulainen}, {Alves}, {Beletsky},
  {Ascenso}, {Kainulainen}, {Amorim}, {Lima}, {Marques}, {Moitinho},
  {Pinh{\~a}o}, {Rebord{\~a}o}, \& {Santos}}]{kainulainen+09}
{Kainulainen}, J.~T., {Alves}, J.~F., {Beletsky}, Y., {et~al.} 2009, \aap, 502,
  L5

\bibitem[{{Kraft} {et~al.}(2001){Kraft}, {Kregenow}, {Forman}, {Jones}, \&
  {Murray}}]{kraft+01}
{Kraft}, R.~P., {Kregenow}, J.~M., {Forman}, W.~R., {Jones}, C., \& {Murray},
  S.~S. 2001, \apj, 560, 675

\bibitem[{{Kudritzki} {et~al.}(1999){Kudritzki}, {M{\'e}ndez}, {Feldmeier},
  {Ciardullo}, {Jacoby}, {Freeman}, {Arnaboldi}, {Capaccioli}, {Gerhard}, \&
  {Ford}}]{kudritzki+99}
{Kudritzki}, R.~P., {M{\'e}ndez}, R.~H., {Feldmeier}, J.~J., {et~al.} 1999, The
  Messenger, 98, 50

\bibitem[{{Malin} {et~al.}(1983){Malin}, {Quinn}, \& {Graham}}]{malin+83}
{Malin}, D.~F., {Quinn}, P.~J., \& {Graham}, J.~A. 1983, \apjl, 272, L5

\bibitem[{{Meaburn} \& {Walsh}(1980)}]{meaburn+walsh80}
{Meaburn}, J. \& {Walsh}, J.~R. 1980, \mnras, 193, 631

\bibitem[{{Merrett} {et~al.}(2006){Merrett}, {Merrifield}, {Douglas},
  {Kuijken}, {Romanowsky}, {Napolitano}, {Arnaboldi}, {Capaccioli}, {Freeman},
  {Gerhard}, {Coccato}, {Carter}, {Evans}, {Wilkinson}, {Halliday}, \&
  {Bridges}}]{merrett+06}
{Merrett}, H.~R., {Merrifield}, M.~R., {Douglas}, N.~G., {et~al.} 2006, \mnras,
  369, 120

\bibitem[{{Minniti} \& {Rejkuba}(2002)}]{minniti+rejkuba02}
{Minniti}, D. \& {Rejkuba}, M. 2002, \apjl, 575, L59

\bibitem[{{Minniti} {et~al.}(2004){Minniti}, {Rejkuba}, {Funes}, \&
  {Akiyama}}]{minniti+04}
{Minniti}, D., {Rejkuba}, M., {Funes}, S.~J., \& {Akiyama}, S. 2004, \apj, 600,
  716

\bibitem[{{Moellenhoff}(1981)}]{moellenhoff81}
{Moellenhoff}, C. 1981, \aap, 99, 341

\bibitem[{{Neumayer}(2010)}]{neumayer10}
{Neumayer}, N. 2010, \pasa, 27, 449

\bibitem[{{Parker} {et~al.}(2012){Parker}, {Frew}, {Acker}, \&
  {Miszalski}}]{parker+12}
{Parker}, Q.~A., {Frew}, D.~J., {Acker}, A., \& {Miszalski}, B. 2012, in IAU
  Symposium, Vol. 283, IAU Symposium, 9--16

\bibitem[{{Pasquini} {et~al.}(2002){Pasquini}, {Avila}, {Blecha}, {Cacciari},
  {Cayatte}, {Colless}, {Damiani}, {de Propris}, {Dekker}, {di Marcantonio},
  {Farrell}, {Gillingham}, {Guinouard}, {Hammer}, {Kaufer}, {Hill}, {Marteaud},
  {Modigliani}, {Mulas}, {North}, {Popovic}, {Rossetti}, {Royer}, {Santin},
  {Schmutzer}, {Simond}, {Vola}, {Waller}, \& {Zoccali}}]{pasquini+02}
{Pasquini}, L., {Avila}, G., {Blecha}, A., {et~al.} 2002, The Messenger, 110, 1

\bibitem[{{Peng} {et~al.}(2004{\natexlab{a}}){Peng}, {Ford}, \&
  {Freeman}}]{peng+04GCcat}
{Peng}, E.~W., {Ford}, H.~C., \& {Freeman}, K.~C. 2004{\natexlab{a}}, \apjs,
  150, 367

\bibitem[{{Peng} {et~al.}(2004{\natexlab{b}}){Peng}, {Ford}, \&
  {Freeman}}]{peng+04GCS}
{Peng}, E.~W., {Ford}, H.~C., \& {Freeman}, K.~C. 2004{\natexlab{b}}, \apj,
  602, 705

\bibitem[{{Peng} {et~al.}(2004{\natexlab{c}}){Peng}, {Ford}, \&
  {Freeman}}]{peng+04PN}
{Peng}, E.~W., {Ford}, H.~C., \& {Freeman}, K.~C. 2004{\natexlab{c}}, \apj,
  602, 685

\bibitem[{{Peng} {et~al.}(2002){Peng}, {Ford}, {Freeman}, \& {White}}]{peng+02}
{Peng}, E.~W., {Ford}, H.~C., {Freeman}, K.~C., \& {White}, R.~L. 2002, \aj,
  124, 3144

\bibitem[{{Phillips}(1981)}]{phillips81}
{Phillips}, M.~M. 1981, \mnras, 197, 659

\bibitem[{Rejkuba(2004)}]{rejkuba04}
Rejkuba, M. 2004, A\&A, 413, 903

\bibitem[{{Rejkuba} {et~al.}(2007){Rejkuba}, {Dubath}, {Minniti}, \&
  {Meylan}}]{rejkuba+07}
{Rejkuba}, M., {Dubath}, P., {Minniti}, D., \& {Meylan}, G. 2007, \aap, 469,
  147

\bibitem[{{Rejkuba} {et~al.}(2005){Rejkuba}, {Greggio}, {Harris}, {Harris}, \&
  {Peng}}]{rejkuba+05}
{Rejkuba}, M., {Greggio}, L., {Harris}, W.~E., {Harris}, G.~L.~H., \& {Peng},
  E.~W. 2005, \apj, 631, 262

\bibitem[{{Rejkuba} \& {Walsh}(2006)}]{rejkuba+walsh06}
{Rejkuba}, M. \& {Walsh}, J.~R. 2006, in Planetary Nebulae Beyond the Milky
  Way, ed. L.~{Stanghellini}, J.~R. {Walsh}, \& N.~G. {Douglas}, 292

\bibitem[{{Richer} {et~al.}(2010){Richer}, {L{\'o}pez},
  {D{\'{\i}}az-M{\'e}ndez}, {Riesgo}, {B{\'a}ez}, {Garc{\'{\i}}a-D{\'{\i}}az},
  {Meaburn}, {Clark}, {Calder{\'o}n Olvera}, {L{\'o}pez Soto}, \& {Toledano
  Rebolo}}]{diaz_men+10}
{Richer}, M.~G., {L{\'o}pez}, J.~A., {D{\'{\i}}az-M{\'e}ndez}, E., {et~al.}
  2010, \rmxaa, 46, 191

\bibitem[{{Sarzi} {et~al.}(2007){Sarzi}, {Bacon}, {Cappellari}, {Davies},
  {Emsellem}, {Falc{\'o}n-Barroso}, {Krajnovi{\'c}}, {Kuntschner}, {McDermid},
  {Peletier}, {de Zeeuw}, \& {van de Ven}}]{sarzi+07}
{Sarzi}, M., {Bacon}, R., {Cappellari}, M., {et~al.} 2007, \nar, 51, 18

\bibitem[{Schlegel {et~al.}(1998)Schlegel, Finkbeiner, \& Davis}]{schlegel+98}
Schlegel, D.~J., Finkbeiner, D.~P., \& Davis, M. 1998, ApJ, 500, 525

\bibitem[{{Tonry} {et~al.}(2001){Tonry}, {Dressler}, {Blakeslee}, {Ajhar},
  {Fletcher}, {Luppino}, {Metzger}, \& {Moore}}]{tonry+01}
{Tonry}, J.~L., {Dressler}, A., {Blakeslee}, J.~P., {et~al.} 2001, \apj, 546,
  681

\bibitem[{{van den Bergh} {et~al.}(1981){van den Bergh}, {Hesser}, \&
  {Harris}}]{VHH81}
{van den Bergh}, S., {Hesser}, J.~E., \& {Harris}, G.~L.~H. 1981, \aj, 86, 24

\bibitem[{{Walsh} {et~al.}(2012){Walsh}, {Jacoby}, {Peletier}, \&
  {Walton}}]{Walsh+12}
{Walsh}, J.~R., {Jacoby}, G.~H., {Peletier}, R.~F., \& {Walton}, N.~A. 2012,
  \aap, 544, A70

\bibitem[{{Walsh} {et~al.}(1999){Walsh}, {Walton}, {Jacoby}, \&
  {Peletier}}]{Walsh+99}
{Walsh}, J.~R., {Walton}, N.~A., {Jacoby}, G.~H., \& {Peletier}, R.~F. 1999,
  \aap, 346, 753

\bibitem[{{Wilkinson} {et~al.}(1986){Wilkinson}, {Sharples}, {Fosbury}, \&
  {Wallace}}]{wilkinson+86}
{Wilkinson}, A., {Sharples}, R.~M., {Fosbury}, R.~A.~E., \& {Wallace}, P.~T.
  1986, \mnras, 218, 297

\bibitem[{{Woodley} {et~al.}(2010{\natexlab{a}}){Woodley}, {G{\'o}mez},
  {Harris}, {Geisler}, \& {Harris}}]{woodley+10b}
{Woodley}, K.~A., {G{\'o}mez}, M., {Harris}, W.~E., {Geisler}, D., \& {Harris},
  G.~L.~H. 2010{\natexlab{a}}, \aj, 139, 1871

\bibitem[{{Woodley} \& {Harris}(2011)}]{woodley+11}
{Woodley}, K.~A. \& {Harris}, W.~E. 2011, \aj, 141, 27

\bibitem[{{Woodley} {et~al.}(2007){Woodley}, {Harris}, {Beasley}, {Peng},
  {Bridges}, {Forbes}, \& {Harris}}]{woodley+07}
{Woodley}, K.~A., {Harris}, W.~E., {Beasley}, M.~A., {et~al.} 2007, \aj, 134,
  494

\bibitem[{{Woodley} {et~al.}(2005){Woodley}, {Harris}, \&
  {Harris}}]{woodley+05}
{Woodley}, K.~A., {Harris}, W.~E., \& {Harris}, G.~L.~H. 2005, \aj, 129, 2654

\bibitem[{{Woodley} {et~al.}(2010{\natexlab{b}}){Woodley}, {Harris}, {Puzia},
  {G{\'o}mez}, {Harris}, \& {Geisler}}]{woodley+10a}
{Woodley}, K.~A., {Harris}, W.~E., {Puzia}, T.~H., {et~al.} 2010{\natexlab{b}},
  \apj, 708, 1335

\bibitem[{{Young} {et~al.}(2013){Young}, {Scott}, {Serra}, {Alatalo}, {Bayet},
  {Blitz}, {Bois}, {Bournaud}, {Bureau}, {Crocker}, {Cappellari}, {Davies},
  {Davis}, {de Zeeuw}, {Duc}, {Emsellem}, {Khochfar}, {Krajnovic},
  {Kuntschner}, {McDermid}, {Morganti}, {Naab}, {Oosterloo}, {Sarzi}, \&
  {Weijmans}}]{young+13}
{Young}, L.~M., {Scott}, N., {Serra}, P., {et~al.} 2013, ArXiv e-prints

\end{thebibliography}

\end{document}